\documentclass[preprint]{aastex}
\usepackage[latin1]{inputenc}
\usepackage{graphicx}
\graphicspath{ {.} }
\usepackage{amsmath}
\usepackage{hyperref}

\DeclareMathOperator*{\argmin}{arg\,min}

\slugcomment{}

\shorttitle{}
\shortauthors{Guennou et al.}

\begin{document}

\title{Can the Differential Emission Measure constrain the timescale of energy deposition in the corona?\\}

\author{C. Guennou\altaffilmark{1} and F. Auch\`ere\altaffilmark{1}}
\affil{Institut d'Astrophysique Spatiale, B\^atiment 121, CNRS/Universit\'e Paris-Sud, 91405 Orsay, France}
\email{chloe.guennou@ias.u-psud.fr}

\and

\author{J.A. Klimchuk\altaffilmark{2}, K. Bocchialini\altaffilmark{1}}
\affil{Solar Physics Laboratory, NASA Goddard Space Flight Center, Greenbelt, Maryland, USA.}

\and

\author{S. Parenti\altaffilmark{3}}
\affil{Royal Observatory of Belgium, 3 Avenue Circulaire, B-1180 Bruxelles, Belgium}

\begin{abstract}
In this paper, the ability of the \textit{Hinode}/EIS instrument to detect radiative signatures of coronal heating is investigated. Recent observational studies of AR cores suggest that both the low and high frequency heating mechanisms are consistent with observations. Distinguishing between these possibilities is important for identifying the physical mechanism(s) of the heating. The Differential Emission Measure (DEM) tool is one diagnostic that allows to make this distinction, through the amplitude of the DEM slope coolward of the coronal peak. It is therefore crucial to understand the uncertainties associated with these measurements. Using proper estimations of the uncertainties involved in the problem of DEM inversion, we derive confidence levels on the observed DEM slope. Results show that the uncertainty in the slope reconstruction strongly depends on the number of lines constraining the slope. Typical uncertainty is estimated to be about $\pm 1.0$, in the more favorable cases.        
\end{abstract}


\section{Motivations}
\label{sec:motivations}

The understanding of how the Sun's outer atmosphere is heated to very high temperatures remains one of the central issues of solar physics today. The physical processes that transfer and dissipate energy into the solar corona remain unidentified and a variety of plausible mechanisms have been proposed \citep[see][for a review of the various coronal heating models]{parnell2012, klimchuk2006, walsh2003, zirker1993}. If the magnetic origin of coronal heating seems to be currently well-accepted~\citep{reale2010}, the details regarding the energy transport from the photosphere to the corona or the energy conversion mechanisms are still open issues. Recently, efforts focused on the determination of the timescale of energy deposition in the solar corona, providing constraint on the properties of the heating mechanisms and allowing for a distinction between steady and impulsive heating scenarios. The nanoflares theory of~\citet{parker1988} for example, is based on the idea that the corona is heated by a series of ubiquitous small and impulsive reconnection events. However, the term nanoflare is now used in a more general way, referring to any impulsive heating event that occurs on small spatial scale, whatever the nature of the mechanism\citep[see][]{cargill1994, cargill2004, klimchuk2001}. Even wave heating takes the form of nanoflares by this definition~\citep[see][]{klimchuk2006}.

According to the impulsive or steady nature of the heating, coronal loops are predicted to present different physical properties at a given time. Observations suggest that coronal loops are probably not spatially resolved. For this reason more often a loop is modeled as a collection of unresolved magnetic strands, considering a strand as a fundamental flux tube with an isothermal cross-section. Depending on the timescale of the heating mechanisms involved, the plasma within the individual strand is allowed or not to cool and drain, \textit{via} a combination of conductive and radiative cooling \citep{reale2010}. Therefore, the thermal structure of the whole loops will differ, the proportion of hot to warm material depending on the time delay between heating events.

Recently, several authors took a particular interest in one potential diagnostic of the heating frequency based on the analysis of the slope of the Differential Emission Measure (DEM) of Active regions (ARs). Based on both theoretical and observational analysis, earlier analysis reported that the coolward part of the DEMs generally follows a power law, up to the emission measure peak ($\sim 3-5$ MK): $\mathrm{DEM}(T) \propto T^{\alpha}$ with $\alpha$ the positive slope index \citep[]{jordan1980, dere1982, brosius1996}. This slope provides indications directly related to the heating timescale: a large proportion of hot relative to warm material leads to a steep DEM slope, whereas a shallower slope corresponds to less hot material and more warm material. The former case is consistent with high frequency impulsive heating, where the short time delay (lower or equivalent to the cooling time) between two heating events does not allow the cooling of a large proportion of material. In the latter case, the time delay between two heating events (now larger than the cooling time) allows the cooling of a significant quantity of the strand material. The limiting case, where the time delay tends to zero, actually corresponds to the steady heating case, where the strand is continuously heated. Using different combinations of observations from the Extreme-ultraviolet Imaging Spectrometer~\citep[EIS;][]{culhane2007} on board the Japanese mission \textit{Hinode}~\citep{kosugi2007}, the Atmospheric Imaging Assembly~\citep[AIA;][]{lemen2012} instrument on board the \textit{Solar Dynamic Observatory} (SDO), and the \textit{Hinode} soft X-Ray Telescope \citep[XRT;][]{golub2007}, several authors recently carried out new AR observational analysis, estimating slope values ranging from 1.7 to 5.17 for 21 different AR cores~\citep{tripathi2011, warren2011, winebarger2011, schmelz2012, warren2012}.   

In the present work, we focus on the investigation of the possibilities to derive the DEM from observations, and we provide a method to estimate the uncertainties associated with its parameters, especially the slope. We do not refer to any particular physical mechanism, such as magnetic reconnection or dissipation of waves, we only refer to the timescale of the mechanism \textit{itself}. Technical difficulties related to both observational processing and diagnosis complicate the slope derivation and thus the associated physical interpretation. In particular, the DEM inversion problem has proved to be a real challenge, due to both its intrinsic underconstraint and the presence of random and systematic errors. Authors were early attentive to examining the fundamental limitations of this inversion problem~\citep{craig1976, brown1991, judge1997}, and many different inversion algorithms have been proposed~\citep{craig1986, landi1997, kashyap1998, mcintosh2000, goryaev2010, hannah2012}. Despite all these attempts, reliably estimating the DEM and the uncertainties associated with the solution remain a major obstacle to properly interpret the observations. 

In this perspective, we developed in recent papers~\citep[][hereafter Paper I and II]{guennou2012a, guennou2012b} a technique, applicable to broadband or spectroscopic instruments, able to completely characterize the robustness of the DEM inversion in specific cases. Using a probabilistic approach to interpret the DEM solution, this technique, briefly recalled in Section~\ref{sec:methodology},  is useful for examining the DEM inversion properties and provides new means of interpreting the DEM solutions. Assuming that the DEM follows a power law, and applying our technique to the \textit{Hinode}/EIS instrument, we derive estimates of the errors associated with the reconstructed DEM slopes, described in Section~\ref{sec:results}. The presence of uncertainties radically changes the conclusions regarding the compatibility between observations and models, as shown by~\citet[see][]{bradshaw2012} and described in Section~\ref{sec:results} and~\ref {sec:conclusions}, where we also discuss the results in the context of steady \textit{vs} impulsive coronal heating .

\section{Methodology}
\label{sec:methodology}

The approach used in this work is very similar to that used in Paper I and II. The technique and the DEM formalism are exhaustively described therein, but a quick summary is given below.   

\subsection{Background}
\label{sec:background}

Under the assumption of an optically thin plasma, the observed intensity in a spectral band $b$ can be expressed as 
\begin{equation}
I_b=\frac{1}{4\pi}\int_{0}^{+\infty}\! R_b(T_e, n_e)\, \xi(T_e)\, \mathrm{d} \log T_e,
\label{eq:int_dem}
\end{equation} 
where $T_e$ is the electron temperature, $\xi(T_e)=\overline{n_e^2}(T_e)\mathrm{d}p/\mathrm{d} \log T_e$ is the DEM\footnote{We choose to define the DEM on a logarithmic scale, but the DEM can also be defined in linear scale as $\xi(T_e)=\overline{n_e^2}(T_e)\mathrm{d}p/\mathrm{d} T_e$. There is a factor $\mathrm{d}\log T_e/\mathrm{d} T_e=1/(\ln 10\ T_e)$ between the two conventions.} that provides a measure of the amount of emitting plasma as a function of temperature, with $\overline{n_{e}^2}$ the square electron density averaged over the portions $dp$ of the Line Of Sight (LOS) at temperature $T_{e}$~\citep{craig1976}. $R_b(T_e)$ is the temperature response function of a given instrument

\begin{equation}
\begin{split}
R_b(n_e, T_e) = & \sum_{X, l} S_b(\lambda_l)\, A_X\, G_{X,l}(n_e, T_e)\\
            & + \int_0^\infty\! S_b(\lambda)\, G_c(n_e, T_e)\, \mathrm{d}\lambda.
\end{split}
\label{eq:response}
\end{equation}   
where the first term refers to the spectral lines $l$ of an atom $X$ of abundance $A_X$, whereas the second describes the contribution of the continua. $S_b(\lambda)$ is the spectral sensitivity of the spectral band $b$ of the instrument, and $G_{X,l}$ and $G_c$ are the contribution functions taking into account all the physics of the coronal emission processes~\citep{mason1994}. The total Emission Measure (EM) is obtained by integrating the DEM over the logarithm of temperature. The inference of the DEM from a set of observations involves the inversion of Equation~\ref{eq:int_dem}, which is hindered by both the presence of random instrumental perturbations and systematic errors on the instrument calibration and on the atomic physics. The purpose of our work here is to investigate the limitations induced by uncertainties in the DEM inversion process concentrating in particular on the determination of the slope of the distribution. Our method is quite general, but we will deal specifically with observations obtained by the \textit{Hinode}/EIS spectrometer. Using simulations of the \textit{Hinode}/EIS observations $I_b^{obs}$ and comparing them to the theoretical expectation $I_b^{th}$, including the perturbations engendered by the uncertainties, it is possible to quantify the reliability of the DEM inversion of the EIS data. 

In simple terms, our approach is essentially the following. We start with an assumed (called "true" hereafter) DEM with a particular functional form. From this we generate a synthetic spectrum, introducing errors associated with unknown atomic physics, instrumental calibration, and photon counting noise. We then determine the DEM that provides the best fit to the synthetic spectrum, which we take to be the DEM that minimizes the differences in the line intensities. This inferred DEM has the same functional form as the true DEM. Only the parameters are different. The most important parameter is the slope, and by comparing the true and inferred slope, we obtain an error in the slope measurement for this particular set of atomic physics, calibration, and noise errors. By running many trials, with many different sets of errors chosen from appropriate probability distributions, we finally deduce an estimate of the uncertainty in the slope determination.

The core of our method resides in the probabilistic approach of the DEM inversion: let us assume a plasma with a true DEM $\xi^T$; the DEM solution $\xi^I$ is the one that minimizes the criterion $C(\xi)$

\begin{equation}
\begin{split}
\xi^I  = & \argmin_{\xi} C(\xi),\\
C(\xi) = & \sum_{b=1}^{N_b} \left( \frac{I_b^{obs}(\xi^T) - I_b^{th}(\xi)}{\sigma_b^u}\right)^{\rm 2}.\\ 
\end{split}
\label{eq:criterion}
\end{equation}    
The solution $\xi^I$ minimizes the distance between the theoretical intensities $I_b^{th}$ and the observed ones $I_b^{obs}$ in $N_b$ spectral bands. The normalization $\sigma_b^u$ corresponds to the standard deviation of the uncertainties. The residuals $\chi^2 = \min C(\xi)$ provide an indication on the goodness of the fit. It is worth noting that as mentioned by~\citet{testa2012a, landi2010}, and Paper I and II, a low $\chi^2$ does not necessarily imply that the solution be the good one or the only one. While our study has broad applicability, we concentrate specifically on observations from the EIS spectrometer on \textit{Hinode}. The criterion is in this case the sum of the contribution of 30 components, one per spectral line. We used the set of 30 lines listed in Table~\ref{tab:eis_lines}, identical to the one used by~\citet{bradshaw2012} and \citet{reep2013} in order to carry out practical comparison between observations and model predictions (see Section~\ref{sec:results}), using the uncertainties derived in this work. Most of them belong to the more prominent lines in the AR regime~\citep{delzanna2005}. Some used lines arise from the same ion species, thus we only have 20 different ion formation temperatures available to constrain the DEM. Column 4 of Table~\ref{tab:eis_lines} indicates the temperatures where the contribution functions peak. However, these additional lines are used in practice as redundant information to decrease the uncertainties. Using Monte-Carlo simulations of the instrumental noises $n_b$ and systematic errors $s_b$ (see Section~\ref{sec:uncertainties} for a detailed description of the uncertainties), the conditional probability $P(\xi^I|\xi^T)$ to obtain the inferred DEM $\xi^I$ knowing that the true DEM is $\xi^T$ can be computed. Then, the inverse conditional probabilities $P(\xi^T|\xi^I)$ giving the probability that the true DEM is $\xi^T$ knowing the inferred results can be deduced from Bayes' theorem. This latter quantity contains all the information possible to extract from a set of observations given the level of uncertainties. 

Thus, the range or multiple ranges of solutions able to explain the observations within the uncertainties can be identified. The derivation of $P(\xi^T|\xi^I)$ requires to know $P(\xi^I)$, and, obviously, because of the uncertainties, a great number of solutions $\xi^I$ can be potentially consistent with a set of observations. Therefore, the computation of this probability is practical only if the space of the solutions is limited, for otherwise it would require the exploration of an infinite number of possible DEMs. For practical reasons, the number of parameters defining the DEM is limited to four: the slope $\alpha$, the temperature of the peak $T_p$, the cutoff at high temperature $\sigma$ and the total EM. 

\subsection{Active Region DEM model}
\label{sec:AR_model}
\begin{figure*}
\begin{center}
\plotone{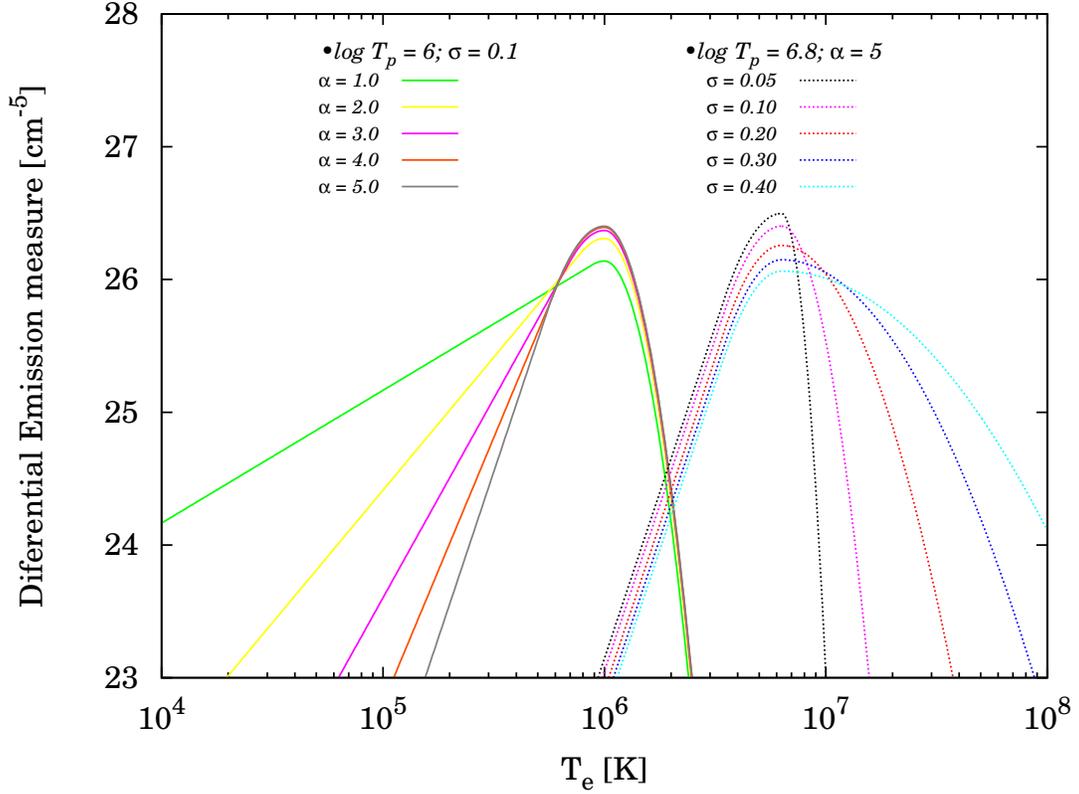}
\end{center}
\caption{Some examples of the parameterization of the AR DEM model (see Section~\ref{sec:AR_model}). The total emission measure is adjusted to the typical AR value of $EM_{AR}=10^{28}$ cm$^{-5}$. The left group illustrates the slope variations, whereas the right group depicts variety of high temperature wing parameterizations. In the first case, the temperature of the coronal peak and the width of the high temperature part are fixed to respectively $T_p = 10^6$~K and $\sigma = 0.1 \log T_e$, while the slope of the five distinct parameterizations varies between 1 and 5. On the right, the peak temperature is increased to $T_p = 10^{6.8}$~K and the slope is fixed to $\alpha = 5$, while the $\sigma$ parameter varies between 0.05 and 0.4.\label{fig:dem_ar}}
\end{figure*}

In order to represent in a more realistic way the observed DEMs, we used the following parameterization of the AR DEM model, represented for different sets of parameters on Figure~\ref{fig:dem_ar}\footnote{For color version of all plots presented in this paper, see the online color version.}. 

\begin{itemize}

\item A power law for the low temperature wing: $T_e < T_0 $
\begin{equation}
\begin{split}
\xi_{AR}(T_{e}) = & k\ \mathrm{EM}\ \times T_e^{\alpha}\\
\mbox{with } k = & T_0^{-\alpha}\ \mathcal{N}_{0.15}(\log T_0 -\log T_{p})\\
\mbox{and } \mathcal{N}_{\sigma}(x) = & \frac{1}{\sigma\sqrt{2\pi}} \exp\left(-\frac{x^2}{2\sigma^2}\right)
\end{split}
\label{eq:dem_low}
\end{equation}
where $\alpha$ is the slope of the DEM coolward of the DEM peak, $T_p$ is the temperature of the DEM peak and EM is the total emission measure. The normalization constant $k$ is used to ensure the continuity and smoothness of the DEM model: the slope must be tangent to the fixed Gaussian connector (see below), at the point $T_0$, depending on the slope value.

\item A Gaussian high temperature wing: $T_e > T_p$
\begin{equation}
\xi_{AR}(T_{e}) =  \mathrm{EM}\ \mathcal{N}_{\sigma}(\log T_e - \log T_p),
\label{eq:dem_high}
\end{equation}
where $\sigma$ is the standard deviation of the Gaussian wing. Thus, beyond the temperature of the DEM peak, the DEM is described by a Gaussian distribution at high temperature, defined by the $\sigma$ parameter. 

\item A fixed width Gaussian connection: $T_0 < T_e < T_{p}$
\begin{equation}
\xi_{AR}(T_{e}) = \mathrm{EM}\ \mathcal{N}_{0.15}(\log T_e - \log T_p),
\label{eq:dem_middle}
\end{equation}
where $T_0$ is the point where the slope $\alpha$ is tangent to the fixed Gaussian $\mathcal{N}_{0.15}$. The connector has been added to ensure that the DEM model is continuous and smooth, corresponding to a continuous first derivative.    

\end{itemize}
A large range of DEM parameters is explored, computing the reference theoretical intensities $I_b^0$, used to deduce $I_b^{obs}$ and $I_b^{th}$ (see Section~\ref{sec:uncertainties}), for electron temperatures $T_e$ ranging from $\log T_e$ = 5 to $\log T_e$ = 7.5 in steps of 0.005 $\log T_e$. The slope $\alpha$ varies from 1.0 to 6.0 in steps of 0.05, and the high temperature wing is explored from $\sigma = 0.01$ to $0.05 \log T_e$ in steps of 0.01. The total EM varies between $3\times 10^{26}$ and $3\times 10^{29}$ cm$^{-5}$ with a resolution of 0.1 in logarithmic scale, and the temperature of the peak $T_p$ varies between $\log T_p = 5.9$ and $\log T_p = 6.9$ in steps of 0.02. Limiting the possible range of each parameter allows us to pre-compute once and for all the reference theoretical intensities $I_b^0$ as a function of the four parameters $\alpha, \sigma, T_p, \mathrm{EM}$, for each of the thirty lines used in this work (Table~\ref{tab:eis_lines}). The variation interval of each parameter is in good agreement with the current observational measurements. Figure~\ref{fig:dem_ar}, illustrates the large range of parameters explored in this work. The emission measure is fixed to the typical AR value of $\mathrm{EM}=10^{28}$ cm$^{-5}$ while the others parameters $\alpha,\ \sigma$ and $T_p$ are allowed to vary. The five curves on the left are all drawn for the same peak temperature $T_p = 10^6$ K and a fixed Gaussian high temperature wing of $\sigma = 0.1 \log T_e$, whereas the slopes varies between 1 and 5. The last five curves on the right display the variation of the high temperature wing: the central temperature $T_p$ and the slope $\alpha$ are now fixed to respectively $T_p = 10^{6.8}$ K and $\alpha = 5$ whereas the $\sigma$ parameter varies between 0.05 and 0.49 .

\subsection{Uncertainties}
\label{sec:uncertainties}
Following the initial reasoning of Paper I, the theoretical intensities $I_b^{th}$ and $I_b^{obs}$ can be expressed as $I_b^{th} = I_b^{0} + s_b$ and $I_b^{obs} = I_b^{0} + n_b$, where $I_b^0$ are called the reference theoretical intensities, $n_b$ are the random perturbations and $s_b$ are systematic errors. The reference theoretical intensities are equal to $I_b^{obs}$ and $I_b^{th}$ in case of a hypothetically perfect knowledge of the atomic physics and observations. They have been computed \textit{via} Equation~\ref{eq:int_dem} and Equation~\ref{eq:response} and using the given AR DEM model $\xi_{AR}$ (see Section~\ref{sec:AR_model}). We used the CHIANTI 7.1 atomic database~\citep{dere1997, landi2013}, and for each of the spectral lines $b$ listed in Table~\ref{tab:eis_lines}, the EIS reference theoretical intensities have been calculated using the function \texttt{eis\_eff\_area}~\citep{mariska2010} of the Interactive Date Language \textit{Solar Software} (SSW) package. 

The different nature of the random and systematic uncertainties $n_b$ and $s_b$ affects the observations in distinct ways~\citep{taylor1997}. The random errors affect the data in an unpredictable way, i.e. they could be revealed by a hypothetically large number of experiments, the error made on each measurements differing for each attempts. A set of \textit{Hinode}/EIS observations is randomly perturbed by various factors: the Poisson photon shot noise and the detection noises, such as thermal or read noise, often assumed to be Gaussian. These phenomenona are well-known and can be realistically simulated: Poisson perturbations $P_{\lambda}$ and $\sigma_{ccd}  =6 e^{-}$ rms~\citep{mcfee2003} of Gaussian CCD read noise are added, before conversion to digital numbers (DNs), using the conversion gains of the EIS spectrometer.

In contrast, the systematic uncertainties can not be revealed by the repetition of the same experience, always pushing the results in the same direction and thus leading to a systematic and \textit{unknown} over or under-estimation. Besides, it is difficult to estimate the probability distribution of the systematics. In the following, the probability distribution of such kind of uncertainties will be considered to be Gaussian, as it generally assumed. The observational intensities $I_b^{obs}$ are affected by the uncertainty associated with the calibration of the instrument, estimated by~\citet{culhane2007} to be around $\sigma_{cal} = 25$\% for the two different CCDs cameras of the EIS instrument. This uncertainty refers to the absolute calibration. We used two independant Gaussian variables to model it, one for each camera. All the lines falling on one camera are perturbated by the same amount for each random realization of the uncertainties. The difference between the two cameras can be as large as 40\%. In the second set of uncertainties described in Section~\ref{sec:results}, this difference is reduced to 20\%. Besides, the degradation of the instrument response over time can also include an additional systematic uncertainty, biasing the results in a given direction. 

The theoretical expectations $I_b^{th}$ are impacted by a complex chain of uncertainties of different nature. Thus, the estimation of the errors on the contribution functions $G_c$ and $G_{X,l}$ (see Equation~\ref{eq:response}) is a more challenging task. In particular, recasting the expression of the observed intensities into Equation~\ref{eq:int_dem} is possible only via several implicit physical assumptions~\citep{judge1997}: the plasma is considered as an optically thin gas, in statistical and ionization equilibrium. The electron velocity distributions function are generally considered to be Maxwellian, as in the CHIANTI database, and the abundance of each element must be constant over the LOS. A discrepancy of the observed coronal plasma with one of these assumptions potentially affects the interpretation of the data. For example, the observed enhancement of the low first ionization potential (FIP) elements~\citep{young2005} in the solar corona possibly induces a non-uniformity of the abundances along the LOS.
  
Incompleteness in the atomic databases, such as missing transitions, or inaccuracy in some physical parameters such as ion-electron collision cross sections, de-excitation rates, etc..., also results in systematic uncertainties. For example, the recent release from version 7.0 to version 7.1 of the CHIANTI spectral code~\citep{landi2013}, including important improvements in the soft X-ray data, clearly shows that the version 7.0 of the CHIANTI database was incomplete in the 50-170 $\AA$ wavelength range, leading to strong inaccuracy in the emissivity calculations of some Fe ions from Fe VIII to Fe XIV. These updates particularly affect the temperature response function of 94 and 335$\AA$ channels of the SDO/AIA instrument. Atomic structure computations are based on two different types of electron scattering calculations: the distorted wave~\citep[see][for details]{crothers2010} or the close coupling approximation~\citep[see][for details]{mccarthy1983}, the latter being generally more accurate. Ionization balance implies equilibrium between the ionization and recombination processes, but if the plasma is out-of-equilibrium or in a dynamic phase, the CHIANTI calculations of line intensities are not consistent with the observations. For example in the case of low frequency heating, the plasma can be out of ionization equilibrium, so that the DEM determination will be incorrect~\citep{sturrock1990}. In that case, temperature-sensitive line ratios of individual ions may be a better way to constrain the models~\citep{raymond1990}. However, these effects should not be important except for very hot plasmas produced by impulsive heating~\citep{bradshaw2011, reale2008}. Within the used temperature range, the evolution is slow enough and the density is high enough that ionization equilibrium is generally a good approximation. Impacts of a deviation of the electron velocity distributions from a Maxwellian on the ionization equilibrium and on the electron excitation rates have been studied by~\citet{dzifcakova1992} and~\citet{dzifcakova2000}, showing that the intensities of spectral lines can be significantly altered. The effects of radiative losses inaccuracy have also been investigated by~\citet{reale2012}, demonstrating that changes in the radiative losses have important impacts on the plasma cooling time, which itself impacts the conclusions of the impulsive heating models. Some studies have been recently carried out to evaluate the impact of using inconsistent atomic physics data in the DEM inversion process~\citep{landi2010, landi2012, testa2012a} and found that the DEM robustness can be significantly altered, leading to important uncertainties on the reconstruction accuracy.

To take into account all these effects, we include the uncertainties in our Monte Carlo simulations using normally distributed random variables. For each realization (each simulation), we choose a number randomly from a Gaussian distribution with a halfwidth $\sigma_i$, considering the four following separate classes:

\begin{itemize}
\item \textit{Class 1}: the first uncertainty class $\sigma_{at}$ involves errors that are different for each and every spectral line, thus we used 30 independent Gaussian random variables to model it (i.e. a different random number for each line). These include errors in the radiative and excitation rates, atomic structure calculations, etc.  
\item \textit{Class 2}: the second class $\sigma_{ion}$ involves errors that are the same for every line of a given ion, but different for different ions. We used the same random number for multiple lines of the same ion (e.g., Fe XIV 264, 270, and 274 $\AA$), but different random numbers for different ions, resulting thus in 20 independent Gaussian random variables (3 different Mg ions, 3 Si ions, 8 Fe ions, 2 S ions and 4 Ca ions). This class corresponds to errors in the ionization and recombination rates.   
\item \textit{Class 3}: the third class $\sigma_{abu}$ involves errors that are the same for every line of a given element, but different for different elements, thus we used 5 different Gaussian variables (one per element). These are errors in the elemental abundances that are unrelated to the first ionization potential (FIP) effect. 
\item \textit{Class 4}: finally, the fourth class $\sigma_{fip}$ involves the additional errors that are the same for every low-FIP elements corresponding to errors on the coronal abundance of such elements. In order to simulate this effect, we adopted a mean FIP bias of 2.5, adding then an uncertainty of $\sigma_{fip}$ on this enhancement factor itself, through an identical Gaussian variable. All our sets of spectral lines, except the two Sulfur lines are finally perturbed in the same way.
\item In addition to these atomic physics uncertainties, a generic uncertainty of $\sigma_{ble} = 15\%$ is added on the blended lines, to account for the added technical difficulties to extract a single line intensity from the data. Blended lines are underlined by a $b$ in the EIS spectral lines list in Table~\ref{tab:eis_lines}. 
\end{itemize}

Each theoretical line intensity $I_b^{th}$, is then modified by the sum of the four random numbers representing the four uncertainty classes (plus a fifth random number in case of blended lines), leading to $I_b^{th} = [(1+R_1)(1+R_2)(1+R_3)(1+R_4, \mathrm{if\ low\ FIP} )(1+R_5, \mathrm{if\ blended})] I_b^0$.  Note that the $R_i$ are equally likely to be positive or negative, and the amplitude of the random number is very likely to be less than the Gaussian halfwidth, but will occasionally be larger and on rare occasion will be much larger.  All the random numbers are reset for each new realization. The resulting uncertainty of each spectral line is reported in column 4, where the $\sigma_{unc}$ is obtained by quadratically summing all the sources of uncertainty, as is appropriate if the errors are independent: $\sigma_{unc}^2 = \sigma_{at}^2 +\sigma_{ion}^2 + \sigma_{abu}^2 + \sigma_{cal}^2 (+ \sigma_{fip}^2 + \sigma_{ble}^2 \ \mathrm{if\ applicable})$.   

In order to determine appropriate amplitudes for the four classes of uncertainty related to atomic physics, we polled a group of well-known solar spectroscopists (G. Del Zanna, G. Doschek, M. Laming, E. Landi, H. Mason, J. Schmelz, P. Young). There was a good consensus that the generic amplitudes are approximately $\sigma_{at} = 20\%$ for the class 1 and $\sigma_{ion} = \sigma_{abu} = \sigma_{fip} = 30\%$ for each of the other three classes. It was noted, however, that the errors could be substantially larger or smaller for specific spectral lines. Adding these uncertainties in quadrature leads to a total atomic physics uncertainty ranging between $46.9$ and $57.6\%$. In subsequent discussions with the spectroscopy experts, the opinion was expressed that a total uncertainty of this magnitude is too large for some well studied lines. Compatibility checks can be applied to observations which sometimes suggest smaller uncertainties. For example, if several lines from the same ion, e.g., Fe XIV, consistently imply a similar EM, then the errors of class 1 (excitation rates) are probably small for those lines. Another example, if the iron lines representing different stages of ionization (Fe X, XI, etc.) follow a consistent trend, such as implying a smooth DEM, then the errors of class 2 (ionization rates) are probably small for these lines. 

We therefore have considered a second set of uncertainties leading to a total uncertainty (i.e. atomic physics plus calibration) ranging between 25 and 30\%, to obtain values of uncertainties typically used in observational analysis: Class 1-4 are now evaluated to 10\%, whereas the calibration errors are decreased to $\sigma_{cal} = 20\%$. The results corresponding to both these sets of uncertainties are presented in Section~\ref{sec:results}. Ultimately, a customized set of uncertainties should be developed for the specific line lists that have been used in published studies. This is beyond the scope of our present investigation, but is something we plan for the future. Until such customized uncertainties are available, it is our opinion that the primary set of uncertainties (20\%, 30\%, 30\%, 30\%) are most appropriate for estimating the uncertainties in the DEM slope. Atomic physics uncertainties are difficult to determine, but the associated systematic errors decreased in the last decades, thanks to more sophisticated computation facilities, and more accurate atomic physic experiments.

Even though we have tried to simulate the systematic errors in a realistic way, some additional sophistications could also be added in our model. Our treatment of the class 1 and 2 uncertainties as intensity modifications is an approximation. In reality, errors in excitation, ionization, and recombination rates are manifested as modifications in the $G_{X,l}$ and $G_c$ contribution functions of the lines (see Equation~\ref{eq:response}). These functions change shape and central position as well as amplitude. A given modification in $G_{X,l}$ or $G_c$ will therefore produce an intensity change that depends on the DEM. Treating this properly could be done in the future but is beyond the scope of this initial work. Future studies might also account for the correlation between various uncertainties. For example, if the class 2 error for Fe XIV is positive, the class 2 error for Fe XIII and Fe XV is likely to be negative.

\section{Results}
\label{sec:results}

In order to quantify the influence of both random and systematic errors, we performed several Monte-Carlo simulations with the uncertainties described in Section~\ref{sec:uncertainties} and the AR DEM model described in Section~\ref{sec:AR_model}. The thirty lines described in Table~\ref{tab:eis_lines} have been used. The simulated observations $I_b^{obs}$ and the theoretical intensities $I_b^{th}$ have been calculated with the same AR DEM model. In this way, the model can perfectly represent the simulated EIS data. 
Since the solutions correspond by definition to the absolute minimum of the least-square criterion (Equation~\ref{eq:criterion}), all solutions are fully consistent with the simulated data. Thus, comparison between the input simulated data and the inversions reveal limitations associated with the presence of uncertainties, and not by the inversion scheme \textit{itself}. We argue that this is actually an optimistic case, since a practical analysis of real observations generally uses blind inversion. The different existing DEM solving algorithms, whether they are based on forward or inverse methods include additional assumptions to ensure uniqueness, such as the smoothness of the solution. Thus, the mathematical difficulties inherent to solving the inverse problem generally introduce additional ambiguity on the results, while our method allows to separate the sources of error and to study the impact of uncertainties only. 

In the following, the four parameters defining the simulated observations with a true AR DEM are denoted $\mathrm{EM}^T$, $T_p^T$, $\sigma^T$ and $\alpha^T$ respectively, whereas the associated inferred parameters resulting from the least-square minimization are noted $\mathrm{EM}^I$, $T_p^I$, $\sigma^I$ and $\alpha^I$. It is useful to think of the coronal plasma parameters as the "true" values, while the inverted one as the "observed" values. To reduce the number of dimensions and for the sake of clarity, we choose to fix the $\mathrm{EM}$ of the simulated observations $I_b^{obs}$ to a constant value $\mathrm{EM}^T_{AR} = 10^{28} \mathrm{cm}^{-5}$, typical of ARs. Since we focus our attention in the ability to reconstruct the slope coolward of the peak of the DEM ($\alpha$ parameter), we also fix the width of the high temperature wing $\sigma$ in both our simulated observations $I_b^{obs}$ and theoretical expectations $I_b^{th}$: only the EM, $\alpha$ and $T_p$ are solved for here. The width $\sigma$ is fixed to the arbitrary constant value $\sigma^T = \sigma^I = 0.2 \log T_e$. We verified that the value of $\sigma$ does not affect results on the slope. Thus, the probability matrices $P(\mathrm{EM}^I, T_p^I, \sigma^I = 0.2, \alpha^I|\mathrm{EM}^T = \mathrm{EM}^T_{AR}, T_p^T, \sigma^T = 0.2, \alpha^T)$ are finally reduced to five dimensions. To illustrate the main properties of these large matrices, we display them by different combinations of fixed parameter values and summation over axes. 

\begin{figure*}
\begin{center}
\plotone{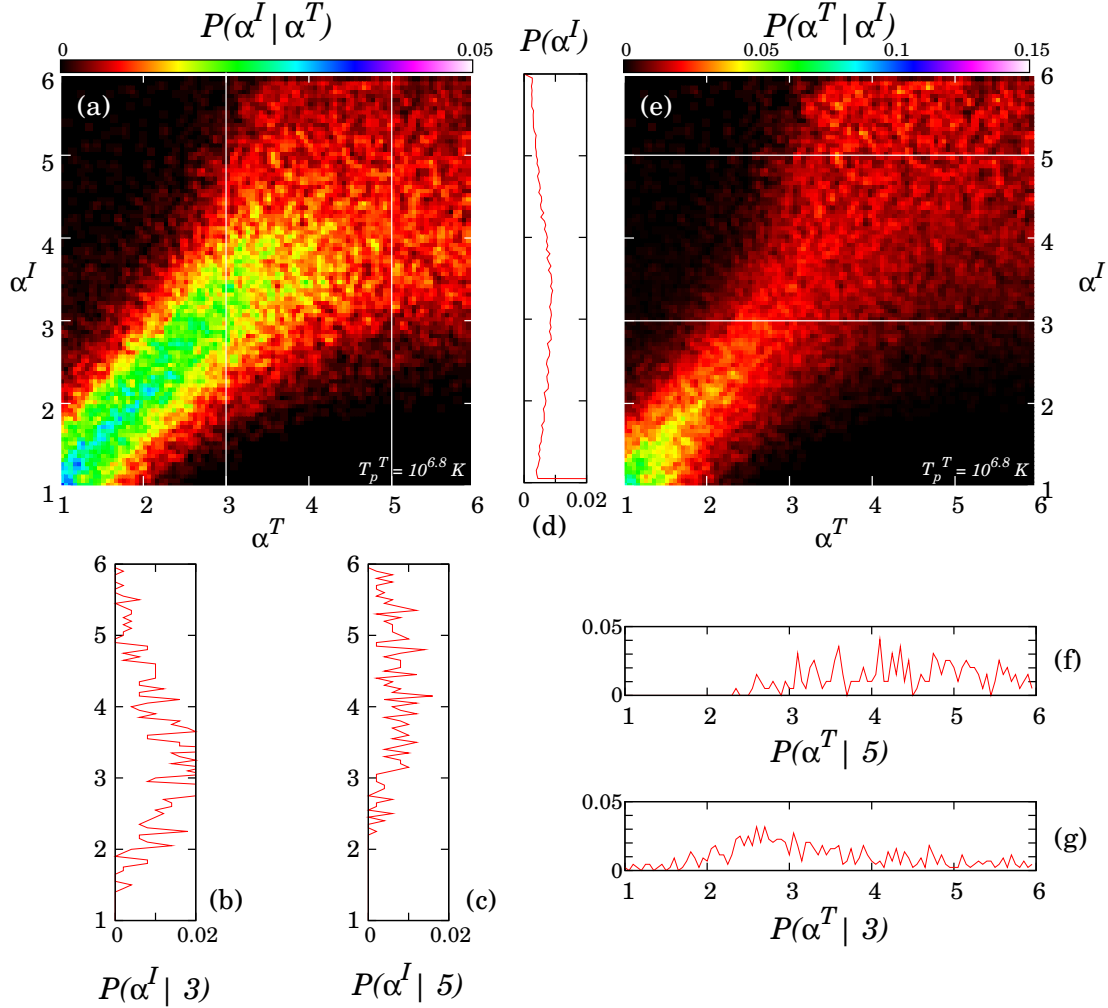}
\end{center}
\caption{Maps of probability for the DEM slope, considering an Active Region (AR) DEM (see Figure~\ref{fig:dem_ar}), and achieved by 1000 Monte-Carlo realizations of the random and systematic errors $n_b$ and $s_b$. In this case, the true DEM is characterized by constant emission measure $\mathrm{EM}_{AR}^T = 10^{28}\ \mathrm{cm}^{-5}$, a fixed high temperature wing of $\sigma^T = 0.2 \log T_e$ and a peak temperature of $T_p^T = 10^{6.8}$~K; only the $\alpha^T$ parameter is investigated here. \textit{(a)}: Probability map $P(\alpha^I|\alpha^T)$, vertically reading. \textit{(b)} and \textit{(c)}: Probability profiles of $\alpha^I$ for true parameter $\alpha^T = 3$ and $5$ corresponding to vertical lines in  panel(a). \textit{(d)}: Total probability $P(\alpha^I)$ to obtain $\alpha^I$ whatever $\alpha^T$. \textit{(e)} \textit{Vice-versa}, probability map $P(\alpha^T|\alpha^I)$, horizontally reading, inferred by means of Bayes' theorem. \textit{(f)} and \textit{(g)}: Probability profiles of $\alpha^T$ knowing that the inversion results are, from top to bottom, $5$ and $3$. From these probability distributions, the slope mean and confidence level are estimated to be $\alpha^T = 4.47 \pm 0.87$ for panel (f) and $\alpha^T = 3.39 \pm 1.07$ for panel (g) (see text in Section~\ref{sec:results} for details).\label{fig:slope_56}}
\end{figure*}

The probability maps resulting from such a simulation are displayed on Figure~\ref{fig:slope_56} for DEMs characterized by a peak temperature of $T_p^T = 10^{6.8}$~K. The probabilities are presented whatever the $\mathrm{EM}^I$ and the peak temperature $T_p^I$ by integrating them over $\mathrm{EM}^I$ and $T_p^I$, even though $\mathrm{EM}^I$ and $T_p^I$ are of course solved for. This allows us to plot two-dimensional probability maps. Panel (a) of Figure~\ref{fig:slope_56} displays the conditional probability $P(\alpha^I|\alpha^T)$\footnote{Defined as the probability for the solutions to be between $\alpha$ and $\alpha + \Delta \alpha$.} of finding a solution $\alpha^I$ knowing the slope $\alpha^T$. Vertically cuts through panel (a) give probability profiles are shown in panels (b) and (c) for the two specific values of $\alpha^T = 3$ and $\alpha^T = 5$ respectively.

The main diagonal structure indicates that the solutions $\alpha^I$ are linearly correlated with the input $\alpha^T$. In $P(\alpha^I|\alpha^T)$ in panel (a) of Figure~\ref{fig:slope_56}, the spreading of the solutions around the diagonal implies that a range of inferred results $\alpha^I$ is consistent with the same true slope parameter $\alpha^T$, given the level of uncertainties involved in this problem. We can also note that for steep slopes, the spreading of the solutions is greater. This is due to the fact that the emission is, in these cases, dominated by higher temperatures, leading to a loss of low temperature lines, which further reduces the temperature range available to constrain the slope. Panels (b) and (c) show ranges of possible inferred solutions for the same true input parameter: considering $\alpha^T = 3$ (panel (b)), the distribution of the solutions $\alpha^I$ is peaked around 3, with more probable values in the $2.5 - 4$ range. In contrast, panel (c) shows that the solutions $\alpha^I$ consistent with the input true slope $\alpha^T = 5$ may be in the 2-6 interval with a quasi uniform distribution. If no additional independent \textit{a priori} information is available, the results of inversion is thus highly uncertain. 

However, the computed probability map $P(\alpha^I|\alpha^T)$ is not usable in a practical way, i.e. with DEM inversion of true observations. Indeed, since the systematics are in reality identical for all measurements, the output $\alpha^I$ will be always biased in the same way. Ignoring to what extent the theoretical intensities are over or under-estimated, we must take into account all the potential inferred solutions. Therefore, in order to deduce the probability distribution of the true parameters $\alpha^T$ consistent with a given inferred result $\alpha^I$ we computed the inverse probability map $P(\alpha^T|\alpha^I)$ using Bayes's theorem (see Section 2.2 of Paper I for more details). This quantity is therefore the relevant one for interpreting a given inferred result $\alpha^I$. Thus, using Bayes' theorem as described in Section~\ref{sec:methodology} and the total probability $P(\alpha^I)$ displayed in panel (d), the inverse conditional probability $P(\alpha^T|\alpha^I)$ shown in panel (e) can be computed. A horizontal cut through panel (e) give the probability distribution of the true slope $\alpha^T$ for a given observed slope $\alpha^I$. Panels (f) and (g) show examples for $\alpha^I = 5$ and $\alpha^I = 3$. The lack of structure in the first case indicates that a large range of true slopes is consistent with the inferred results: $ 3 < \alpha^T < 6$. In the second case, the most likely value of the true slope is similar to the observed slope of 3, but there is again a wide range of true slopes that are consistent with this observed slope. 

The probability distribution of panel (e) is very useful to assist the DEM inversion interpretation: from this we can compute descriptive statistic quantities such as the standard deviation and the mean of the probability distribution for a given $\alpha^I$, which give a quantitative representation of the reconstruction quality and uncertainty. From panel (f), we derived a mean value of $\overline{\alpha}^P = 4.47$ for a given result of $\alpha^I = 5$. The standard deviation, evaluated to 0.87 in this case, characterizing the dispersion of the results, is an estimation of the confidence level on the slope reconstruction. From this, a proper interpretation of the DEM inversion result can be derived, providing a final result of $\alpha^T = 4.47 \pm 0.87$, for a given inferred result of $\alpha^I = 5$. In panel (g), the mean value is estimated to be $\overline{\alpha}^P = 3.39$, whereas the inferred slope was $\alpha^I = 3$. The associated standard deviation is 1.07, leading to a final result of $\alpha^T = 3.39 \pm 1.07$. 

\begin{figure*}
\begin{center}
\plotone{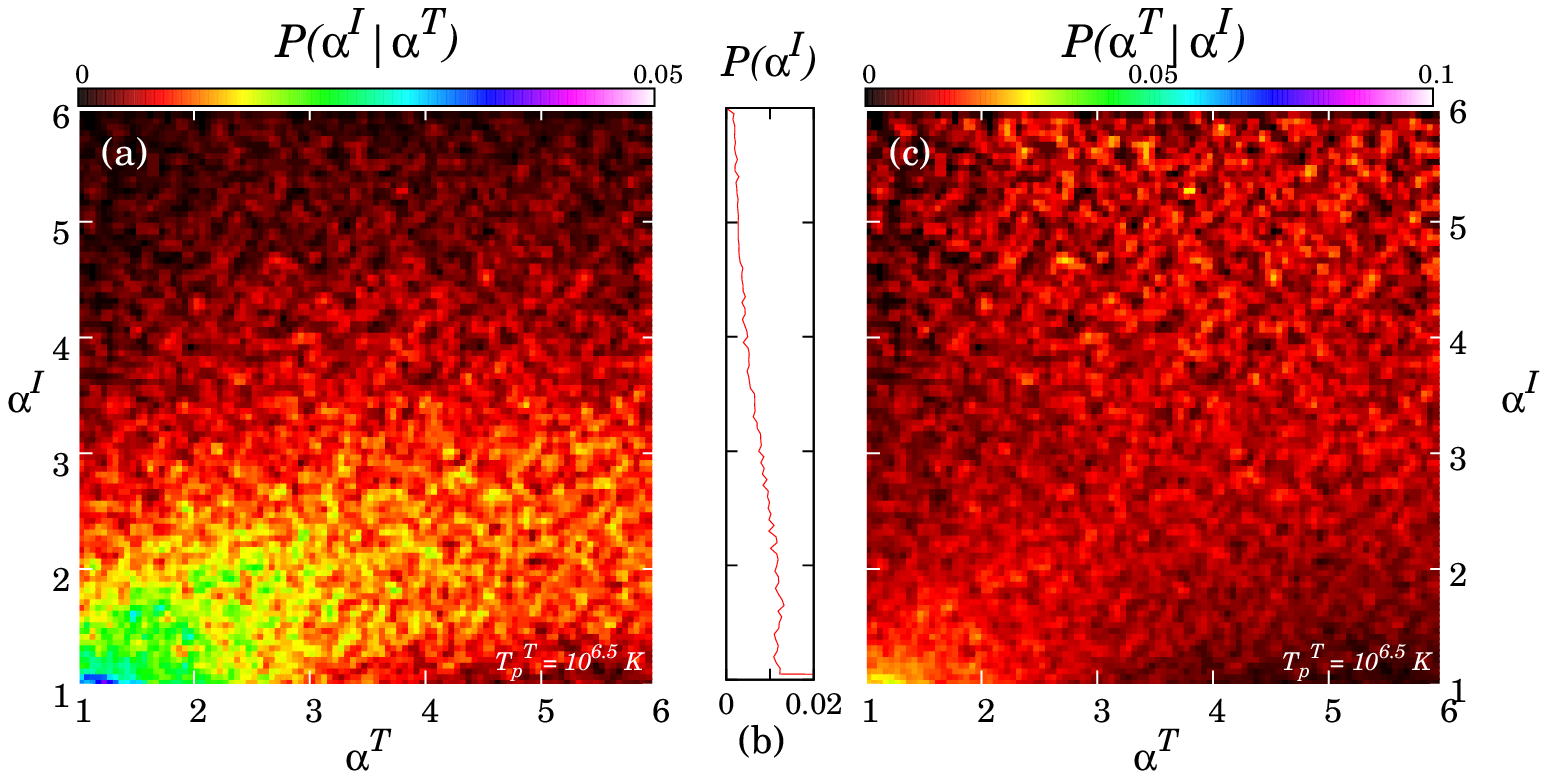}
\plotone{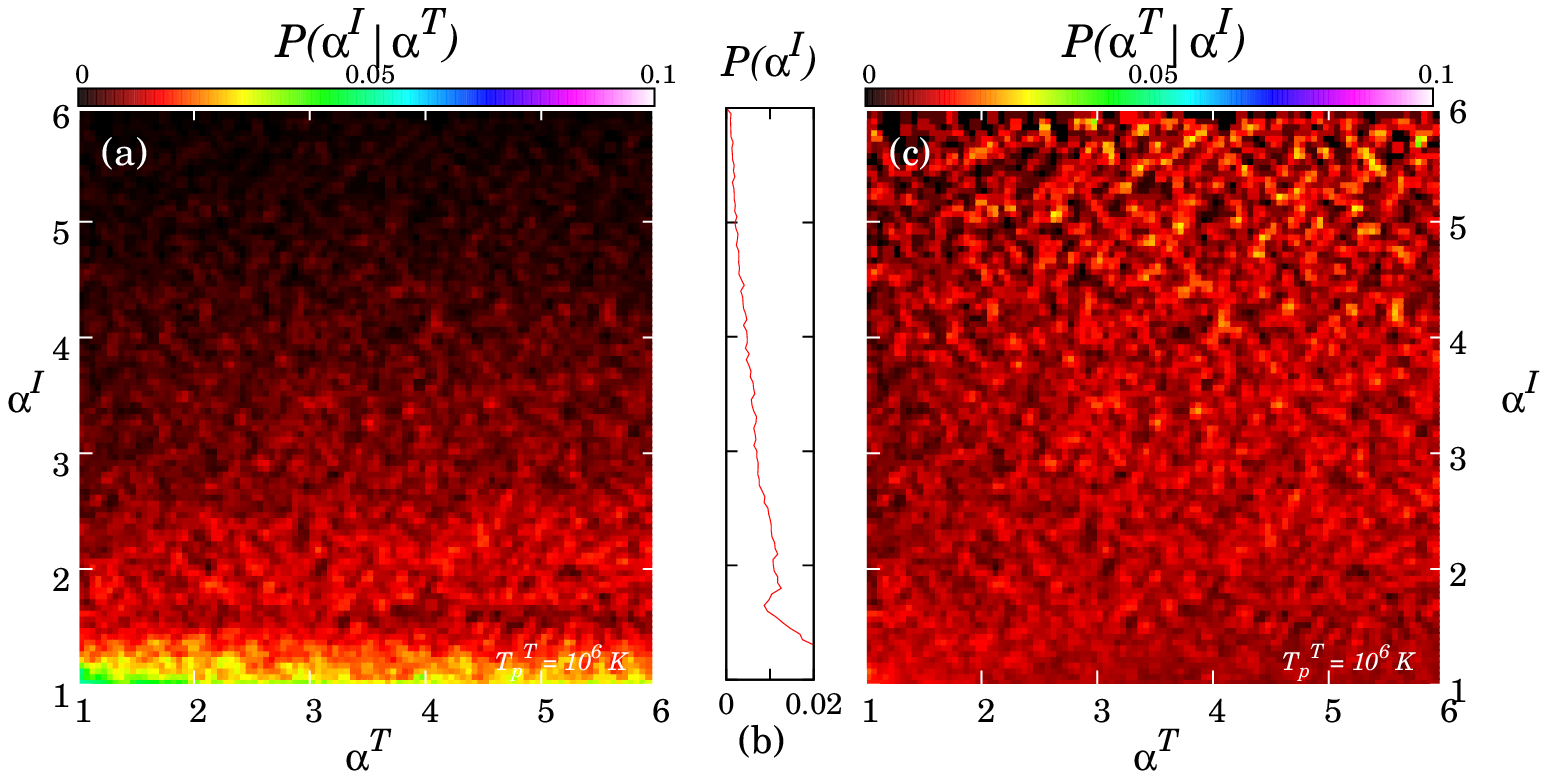}
\end{center}
\caption{Same as Figure~\ref{fig:slope_56}, but with a true DEM characterized by peak temperatures of respectively $T_p^T = 10^{6.5}$ and $T_p^T = 10^{6}$~K, from top to bottom. The decrease of number of lines constraining associated with the uncertainties clearly deteriorate the quality of the inversion, increasing the confidence level to typical value of 1.3 (see also Figure~\ref{fig:statistical}). \label{fig:slope_6}}
\end{figure*}

The situation clearly deteriorates as the temperature of the peak temperature decreases. This is illustrated in Figure~\ref{fig:slope_6} which is the same as Figure~\ref{fig:slope_56}, but now for plasmas with true peak temperatures $T_p^T = 10^{6.5}$ (top) and $T_p^T = 10^{6}$~K (bottom). Compared to the previous case, the probability distributions are clearly wider and less regular. Whatever the inferred result $\alpha^I$, the probability distribution of the possible true solutions $\alpha^T$ extends over the entire possible range. For $T_p^T = 10^{6.5}$~K, we found a typical standard deviation of 1.3-1.4, similar to the one computed for the extreme low temperature peak of $10^{6}$~K. For completeness, the probability maps for 63 peak temperatures $T_p^T$, from $10^{5.9}$ to $10^{6.9}$~K, and an animation showing the whole amplification of the perturbations are available on-line at \texttt{ftp:$\slash\slash$ftp.ias.u-psud.fr$\slash$cguennou$\slash$DEM\_EIS\_inversion$\slash$slope$\slash$slope$\slash$}. This deterioration can be explained by the cumulative effects of the decreasing of number of EIS lines and the smaller temperature range available to constrain the slope part of the DEM. In Figure~\ref{fig:slope_56}, the DEM temperature peak is $T_p^T = 10^{6.8}$~K and thus, all 30 lines constrain the slope and the temperature range in which the slope is allow to vary covers $1.35$ decades. Considering the case displayed on the top of Figure~\ref{fig:slope_6}, where $T_p^T = 10^{6.5}$~K, this number of lines decreases to 26, whereas the temperature range decreases to about 1 decade. In the extreme case of $T_p^T = 10^{6}$~K, only 8 lines constrain the DEM slope, while the temperature range is reduced to only 0.35 decades. 

\begin{figure*}
\begin{center}
\epsscale{0.8}
\plotone{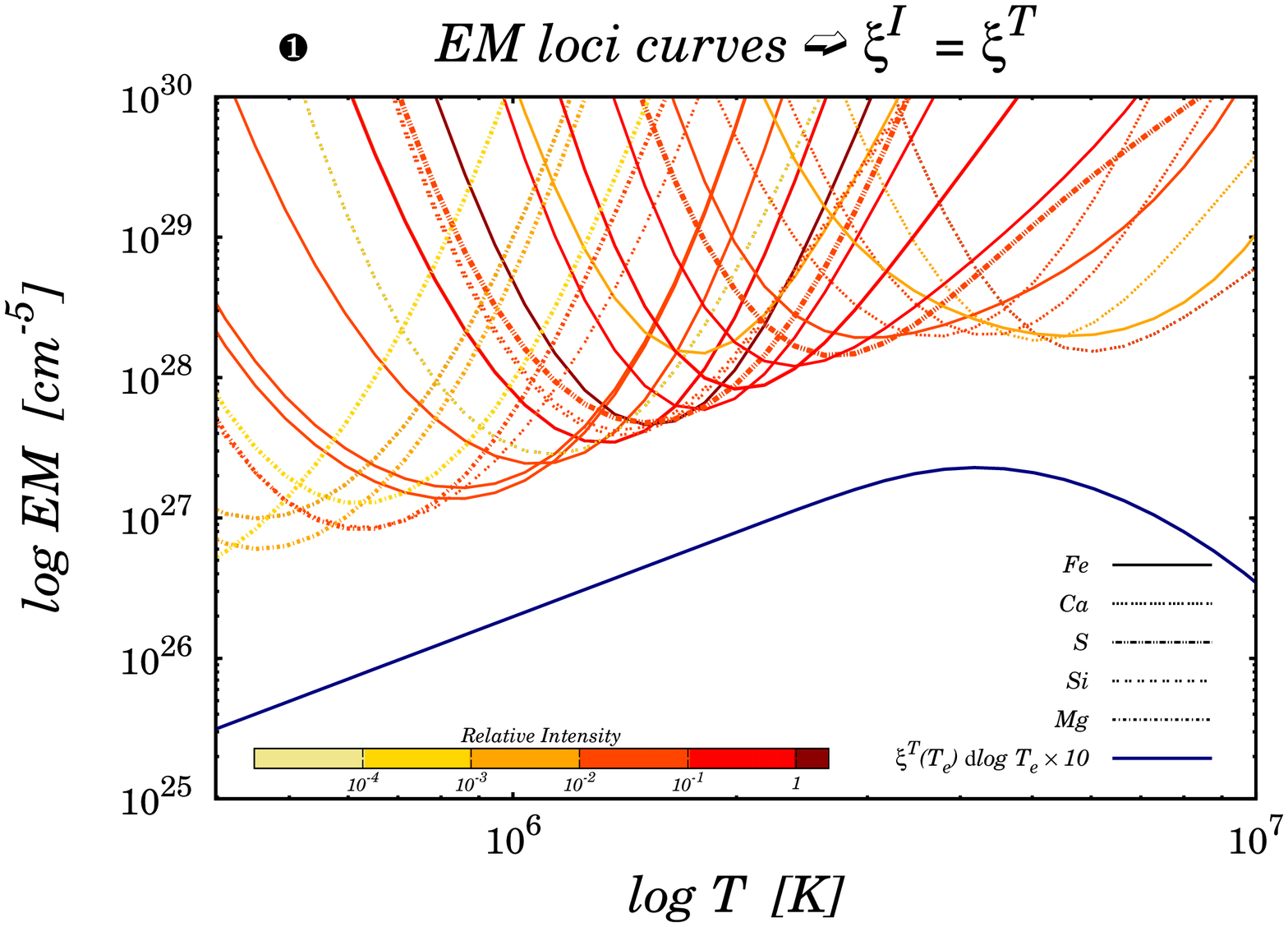}\\%
\plotone{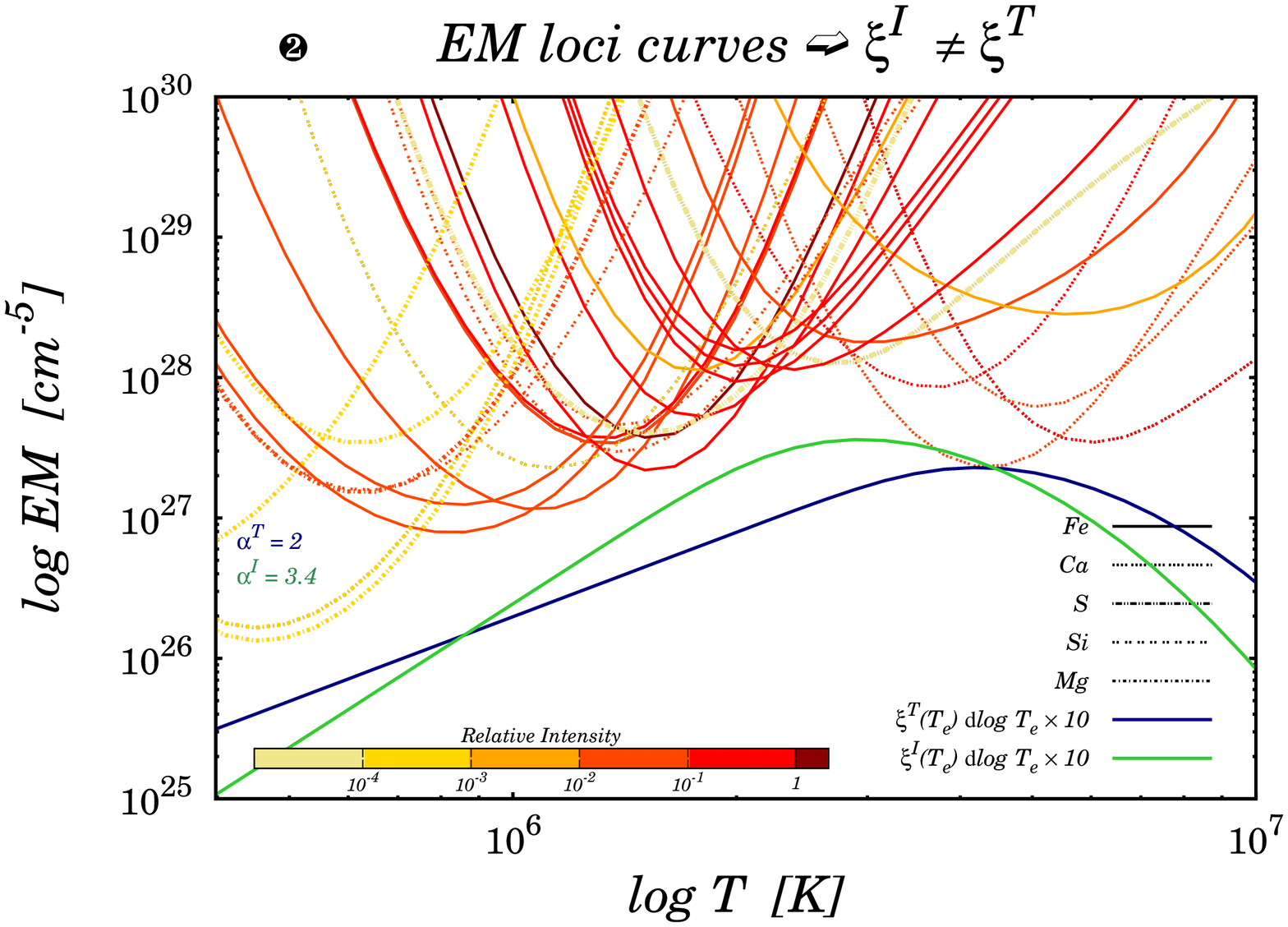}\\
\end{center}
\caption{Illustration of the potential discrepancy between the true DEM $\xi^T$ (blue solid line) and the estimated one $\xi^I$ (green solid line), due to the presence of both random and systematic errors. The EM loci curves are represented as a function of the elements, sorted by line type, and as a function of their relative intensity, sorted by color, from pale yellow (faintest) to dark red (strongest). \textit{Top:} no uncertainty in this first case, thus the inferred DEM $\xi^I$ is equal to the initial one $\xi^T$. \textit{Bottom:} A given realization of systematic and random errors, leading to discrepancy between true and inferred DEMs (see Fig 4).\label{fig:em_loci}}
\end{figure*}

\begin{figure*}
\begin{center}
\epsscale{0.8}
\plotone{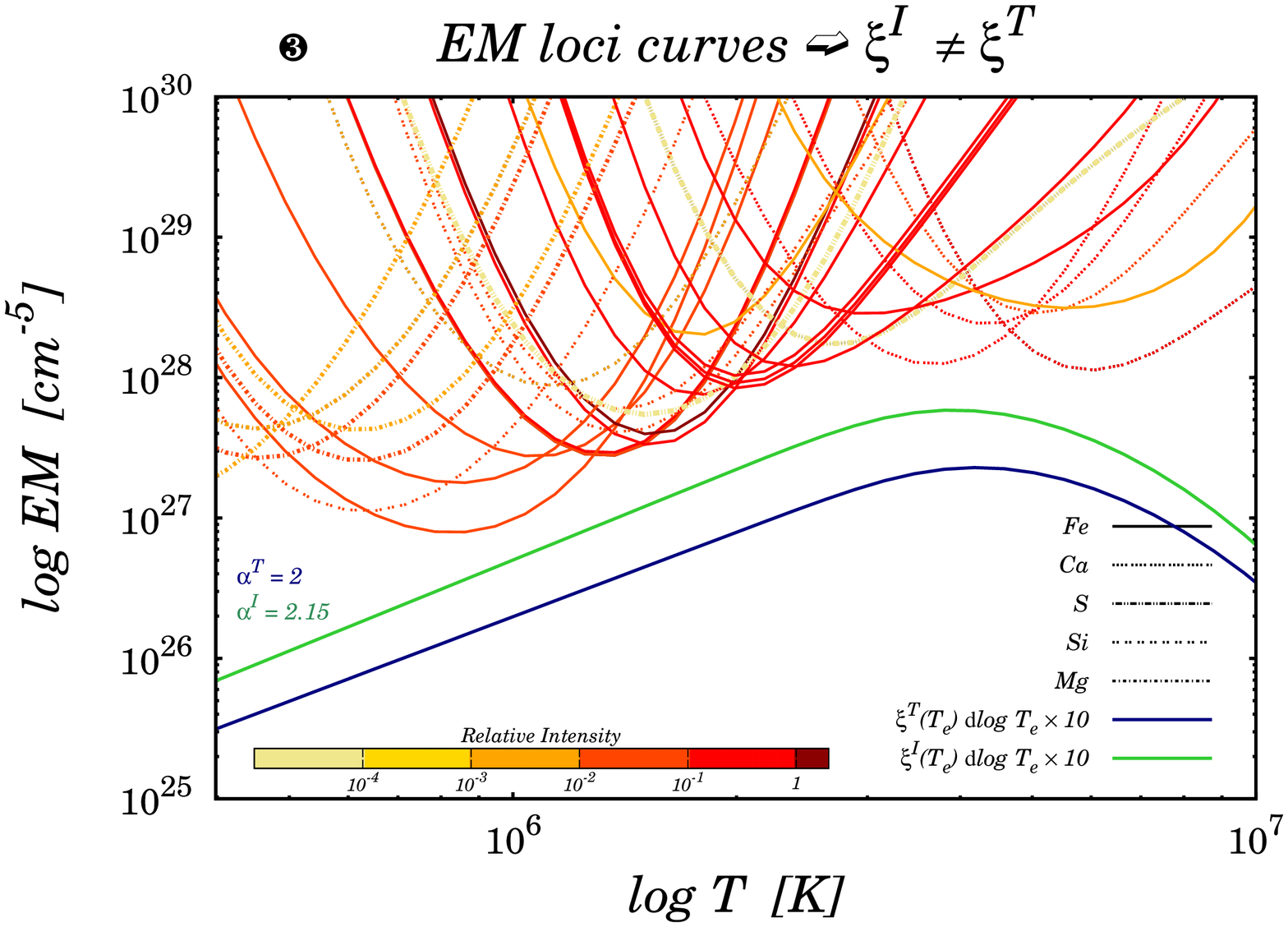}\\
\plotone{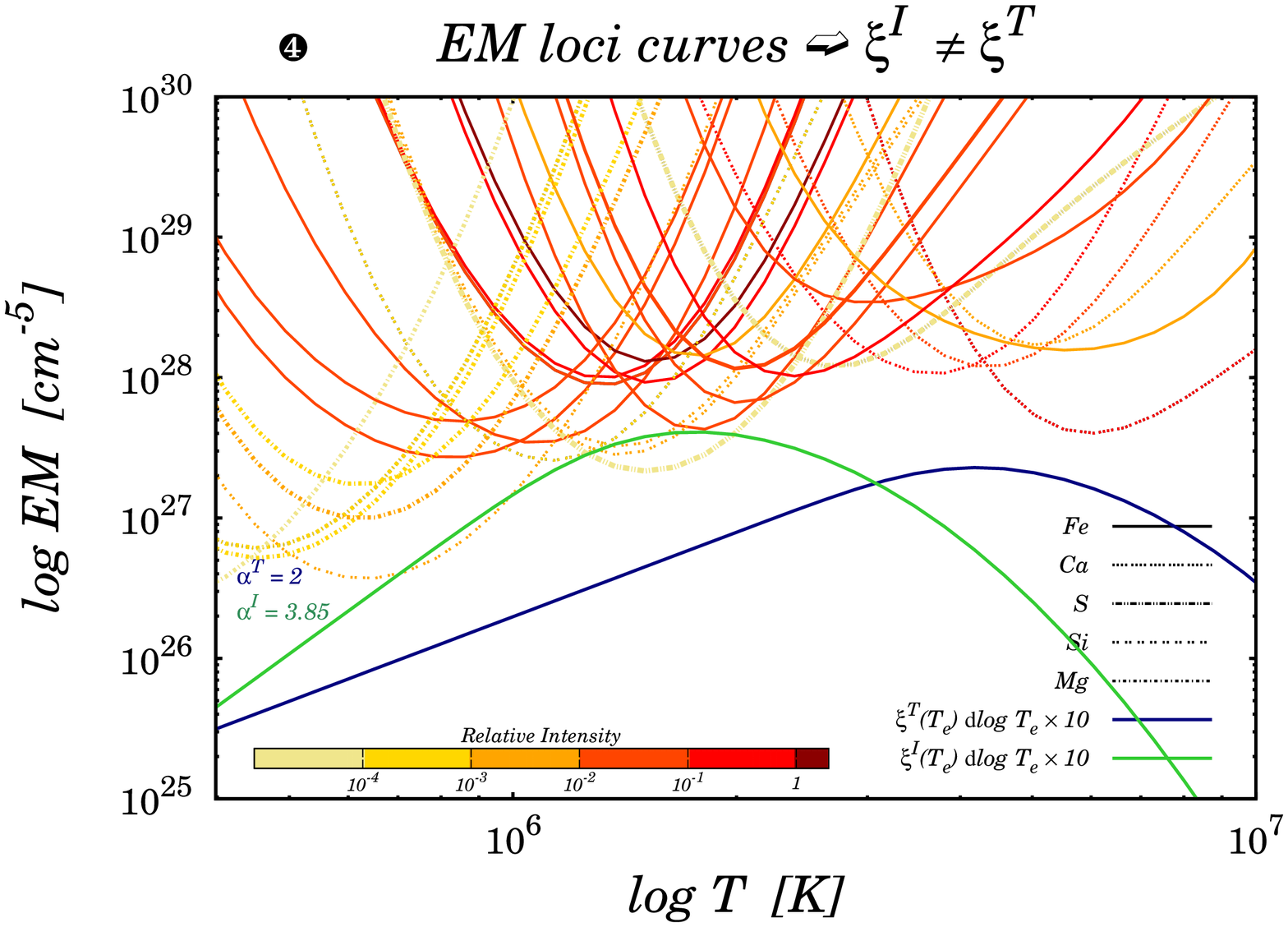}%
\end{center}
\caption{Same as Figure~\ref{fig:em_loci} but for two different realizations of systematic and random errors. The bottom case illustrates an extreme case, leading to strong discrepancy between input and inferred DEMs.\label{fig:em_loci_2}}
\end{figure*}

The potential discrepancy between the true DEM $\xi^T$ and the inferred one $\xi^I$ is illustrated in Figure~\ref{fig:em_loci} and ~\ref{fig:em_loci_2}, by showing three different realizations of uncertainties (Fig.~\ref{fig:em_loci} -bottom- and Fig.~\ref{fig:em_loci_2}), as well as the perfect case (Fig.~\ref{fig:em_loci} -top-). The EM loci curves, formed by the set of (EM, $T_e$) pairs for which the isothermal theoretical intensities exactly match the observations for a given spectral line \citep[see][for more details]{delzanna2003}, are represented for each case as a function of both the element, given by the line type, and the relative intensity, given by the color from pale yellow (faintest) to dark red (strongest). In the case 1 of Figure~\ref{fig:em_loci}, the loci curves are perfectly aligned, and thus the estimated DEM $\xi^T$ perfectly match the initial true DEM $\xi^T$. The case 2 (Figure~\ref{fig:em_loci}) shows a realization of the perturbation $n_b$ and $s_b$, each loci curves being randomly shifted from its original position. This corresponds to a deviation of the solution $\xi^I$, the estimated temperature peak being underestimated from $T_p^T =$4~MK to ~$T_p^I =$2.8~MK and the slope increased to the steeper value of $\alpha^I =$3.4 while the initial true slope was $\alpha^T = 2.0$. Note that the relative intensity of each line plays a key role in the reconstruction: the more intense lines have more important weight in the inversion process, even though we normalize the $\chi^2$ by the different uncertainties sources, including the photon noise (see Equation~\ref{eq:criterion}). The cases 3 and 4 in Figure~\ref{fig:em_loci_2} show another different realizations of errors leading in the case 3 to an overestimation of the total EM, and in the case 4 to a significant deviation of the peak temperature $T_p$.  

\begin{figure*}
\begin{center}
\epsscale{1}
\plotone{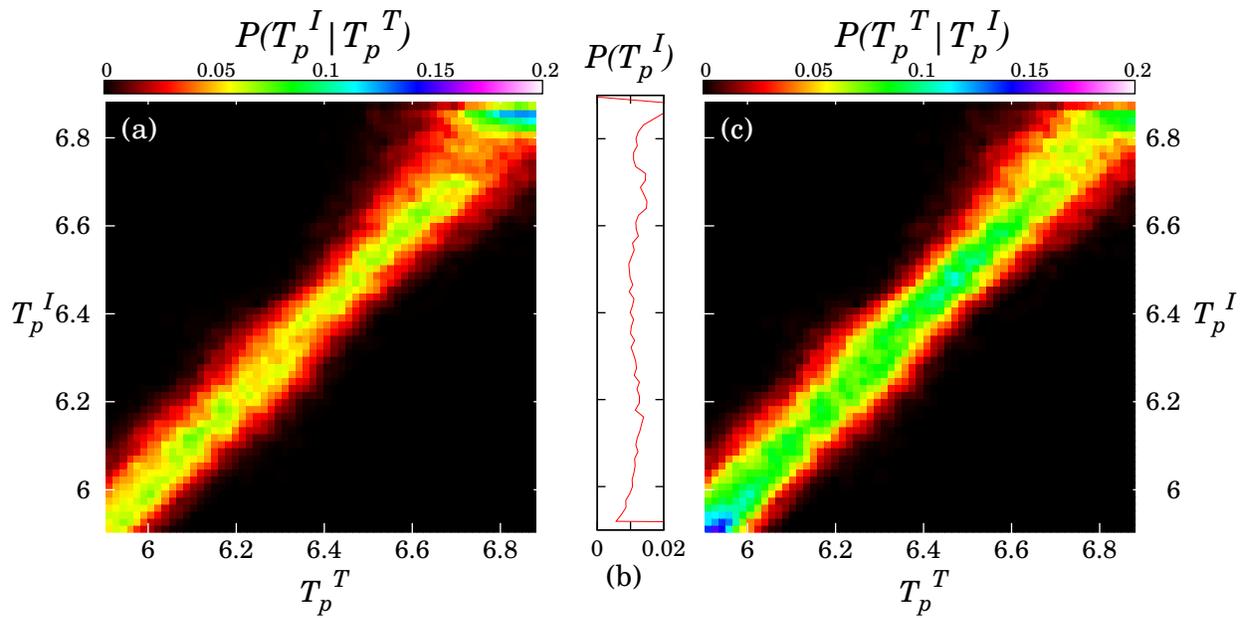}
\end{center}
\caption{Maps of probability for the peak temperature, represented for a simulated observation with a true DEM slope of $\alpha^T= 1.5$. Results originate from same simulations framework than Figure~\ref{fig:slope_56}, showing that if the slope is strongly impacted by the presence of uncertainties, the peak temperature is still well constrained, providing confidence levels between 0.7 and 0.85~MK.\label{fig:temp_slope_6}}
\end{figure*}

The reconstruction of the temperature peak is much better constrained than the slope. Figure~\ref{fig:temp_slope_6} displays the probability maps associated to the $T_p$ parameter, for a true shallow slope $\alpha^T = 1.5$, and a constant AR emission measure $\mathrm{EM}_{AR}^T$. Probabilities are now represented whatever the $\mathrm{EM}^I$ and $\alpha^I$ by integrating them over the $EM^I$ and $\alpha^I$ axes. Results are very similar whatever the chosen input $\alpha^T$, and the probability maps presented here are typical\footnote{The probability maps of the peak temperature for 101 values of $\alpha^T$ ranging from 1 to 6 are available on-line at ftp:$\slash\slash$ftp.ias.u-psud.fr$\slash$cguennou$\slash$DEM\_EIS\_inversion$\slash$slope$\slash$temperature$\slash$}. Most of the solutions are condensed around the diagonal. The use of the thirty lines provides an unambiguous determination of the peak temperature. However, the confidence interval remains quite large: we found a typical standard deviation between 0.7 and 0.85~MK associated to the spread of the solutions around the diagonal for the different tested plasma slopes, with extreme values varying between 0.1 and 1.3~MK.

\begin{figure*}
\begin{center}
\includegraphics[width=0.48\textwidth]{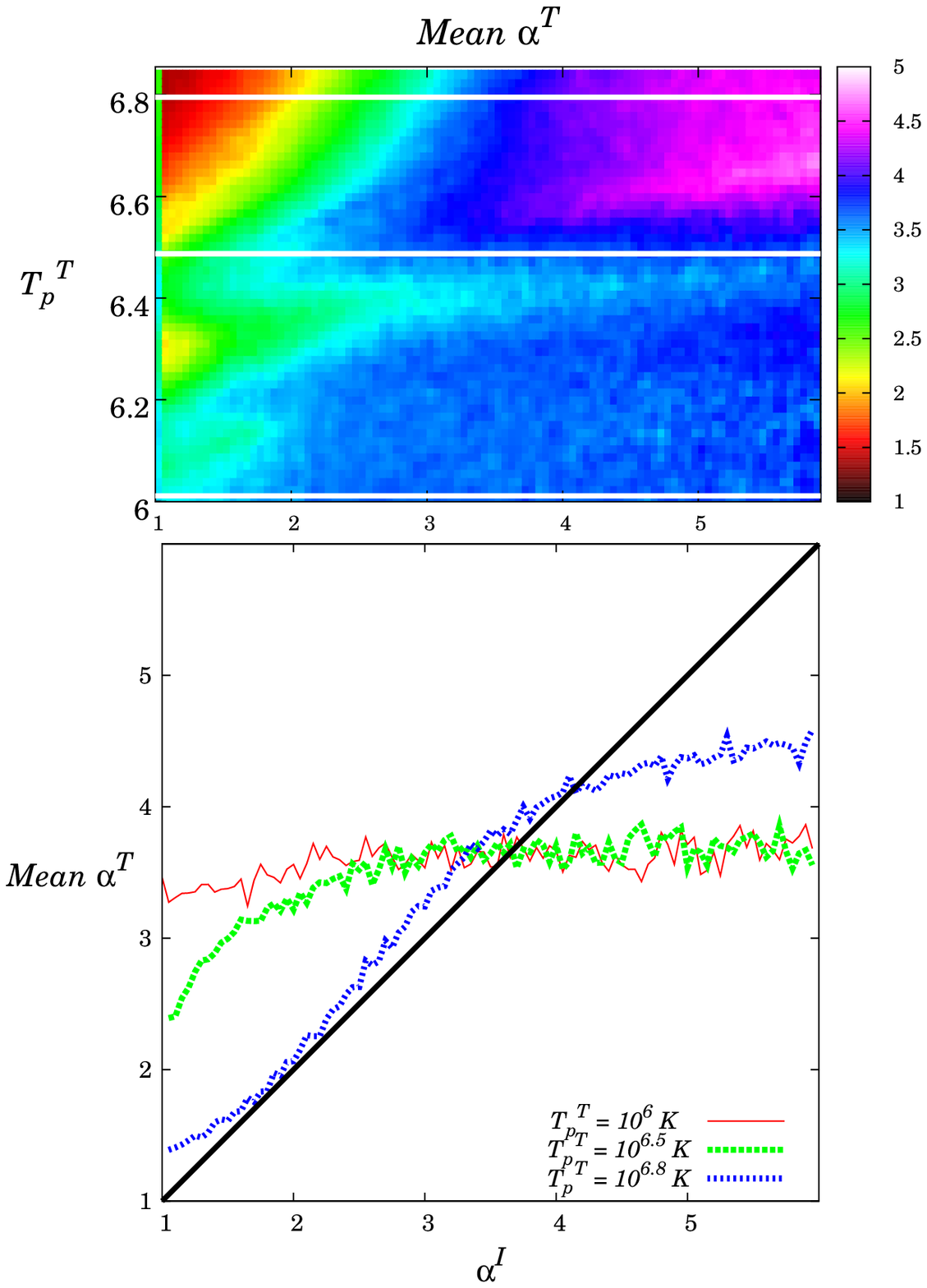}%
\hspace{0.1cm}%
\includegraphics[width=0.48\textwidth]{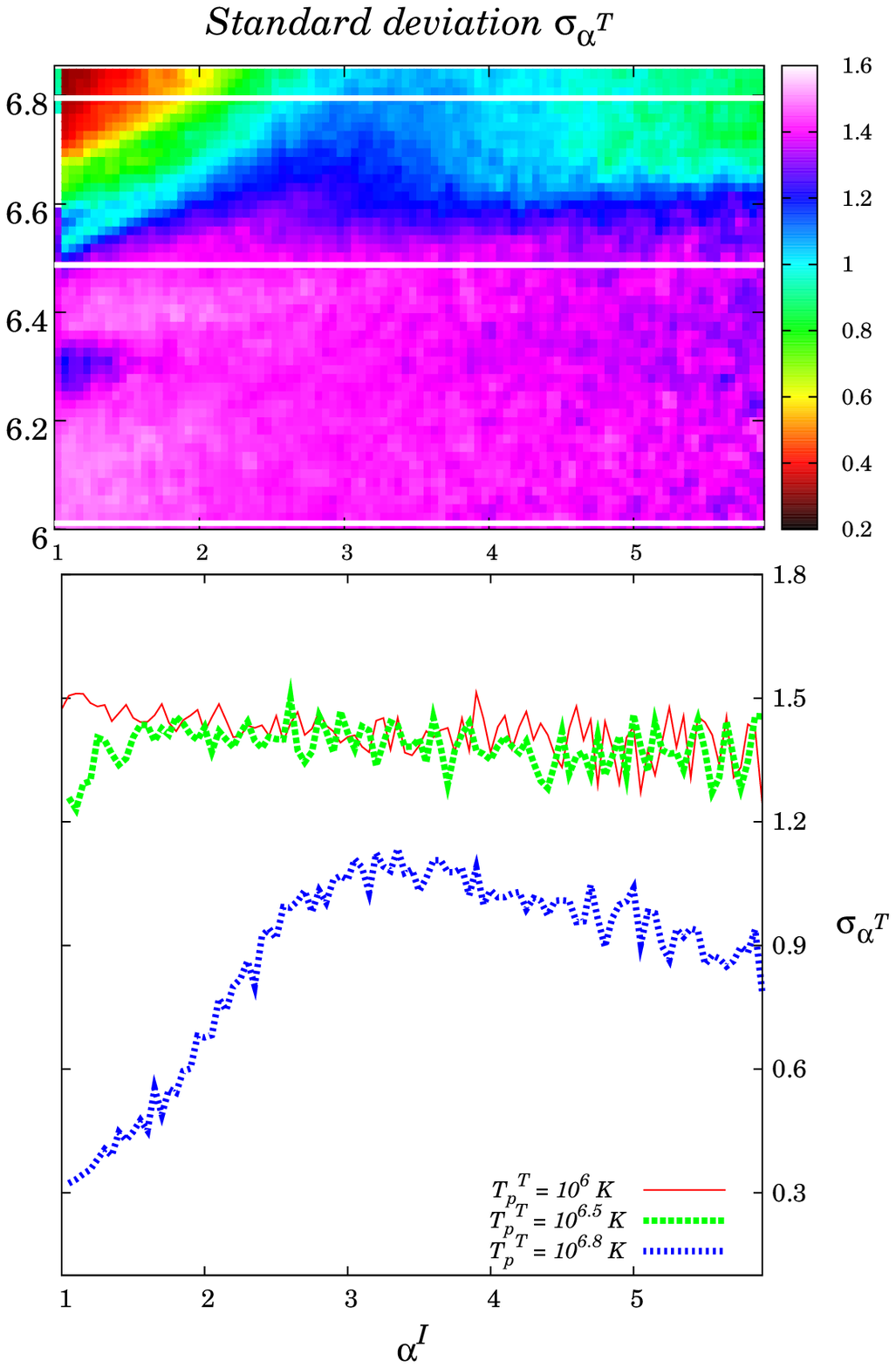}%
\end{center}
\caption{Mean and standard deviation of the true slopes $\alpha^T$ consistent with a given inversion result $\alpha^I$. \textit{Top}: Mean (left) and standard deviation (right) maps represented as a function of the peak temperature and the inversion result $\alpha^I$. \textit{Bottom}: Cut across the mean (left) and standard deviation (right), corresponding to the white horizontal lines. The peak temperature are fixed to respectively $T_p^T = 10^6$~K (solid lines), $T_p^T = 10^{6.5}$~K (bold dashed lines) and $T_p^T = 10^{6.8}$~K (dashed lines), corresponding to the probability maps displayed in Figures~\ref{fig:slope_56} and~\ref{fig:slope_6}. \label{fig:statistical}}
\end{figure*}

These results can finally be summarized in the two graphs of Figure~\ref{fig:statistical}. The first one, on the right, displays the mean slope value of the initial true $\overline{\alpha^T}$, knowing the inferred result $\alpha^I$. On the top, the map shows the slope mean value, represented as a function of both the peak temperature $T_p^T$ and the inferred results $\alpha^I$. The quantity $\overline{\alpha^T}$ has been computed from the probability distribution $P(\alpha^T|\alpha^I)$, in the same way than described previously. The three different horizontal profiles displayed on the bottom and denoted by the horizontal white lines on the top, correspond to the three different probability maps displayed on Figure~\ref{fig:slope_56} and Figure~\ref{fig:slope_6}. Using these curves, it is possible to correctly interpret the results of the inferred $\alpha^I$, providing thus the slope mean value computed from the probability distribution of all true slopes consistent with a given inferred results. The diagonal (black solid line) correspond to a perfect agreement between $\overline{\alpha^T}$ and $\alpha^I$. The bias of $\overline{\alpha^T}$ strongly affects the results for the low temperature profiles $T_p^T = 1$~MK (red solid line) and $T_p^T = 3.2$~MK (green solid line), and in a less significant way the high temperature profile $T_p^T = 6.3$~MK (blue solid line). This bias around the diagonal reflects in reality the initial bias of the solutions observed in the probability maps $P(\alpha^I|\alpha^T)$ previously presented, and taken into account by computing the inverse probability maps $P(\alpha^I|\alpha^T)$. For low temperature peaks, the corresponding probability distributions are very wide, almost covering the whole space of the solutions(see Figure~\ref{fig:slope_6}). Consequently, the slope mean value approaches a roughly constant value of $\overline{\alpha^T} = 3.5$, with, in this case, large associated standard deviation. The behavior of this latter quantity (i.e. the confidence level) is shown on the right side of Figure~\ref{fig:statistical}, uniformly ranging between $\sigma_{\alpha^T} = 1.3-1.4$ for temperature peak lower than $T_p^T = 10^{6.5}$~K, as expected in light of the above. For the high temperature peak $T_p^T = 10^{6.8}$~K, confidence level extends between 0.3 and 1.15, depending on the value of the inferred slope $\alpha^I$.   

\begin{figure*}
\begin{center}
\includegraphics[width=0.48\textwidth]{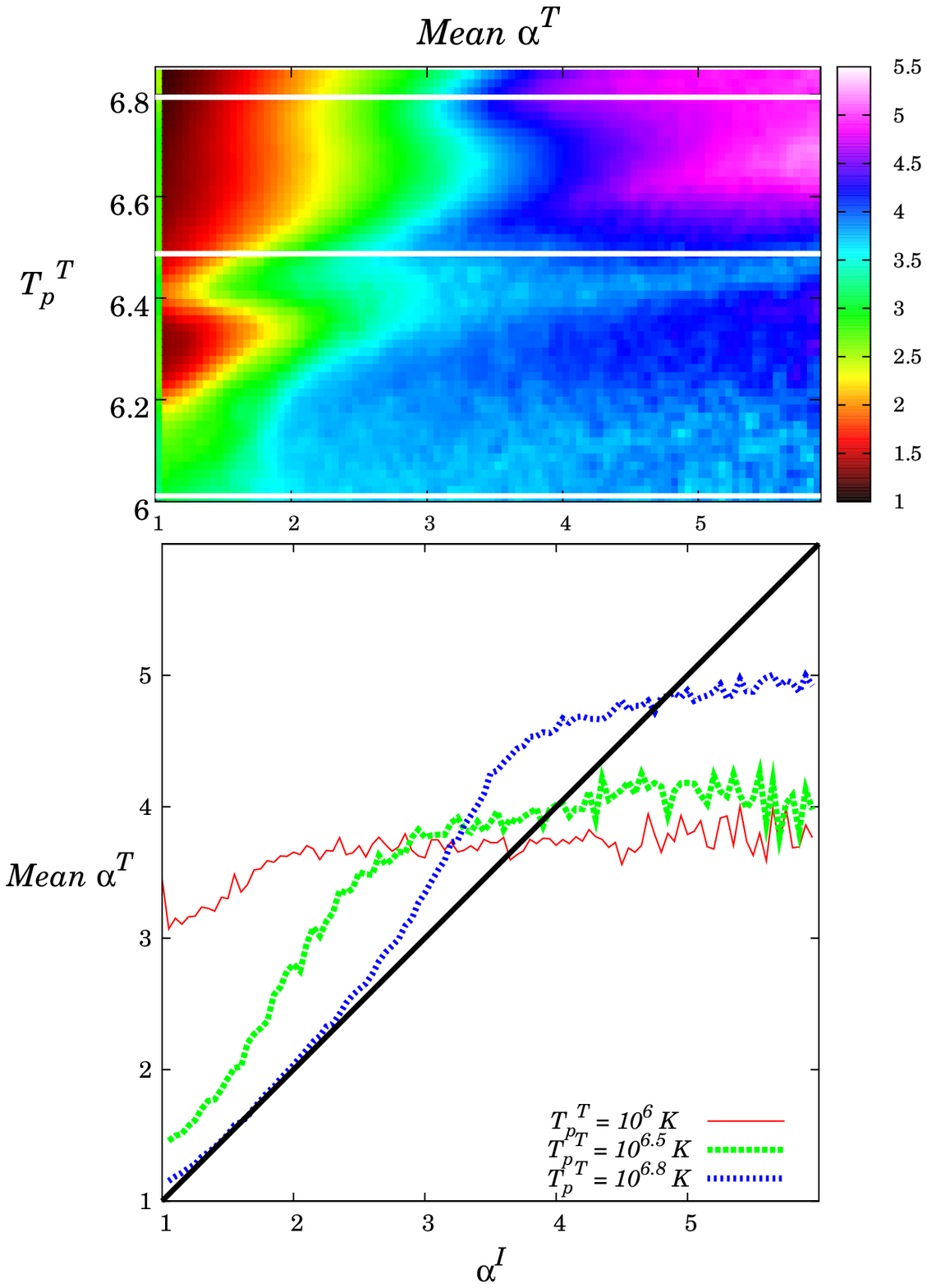}%
\hspace{0.1cm}%
\includegraphics[width=0.48\textwidth]{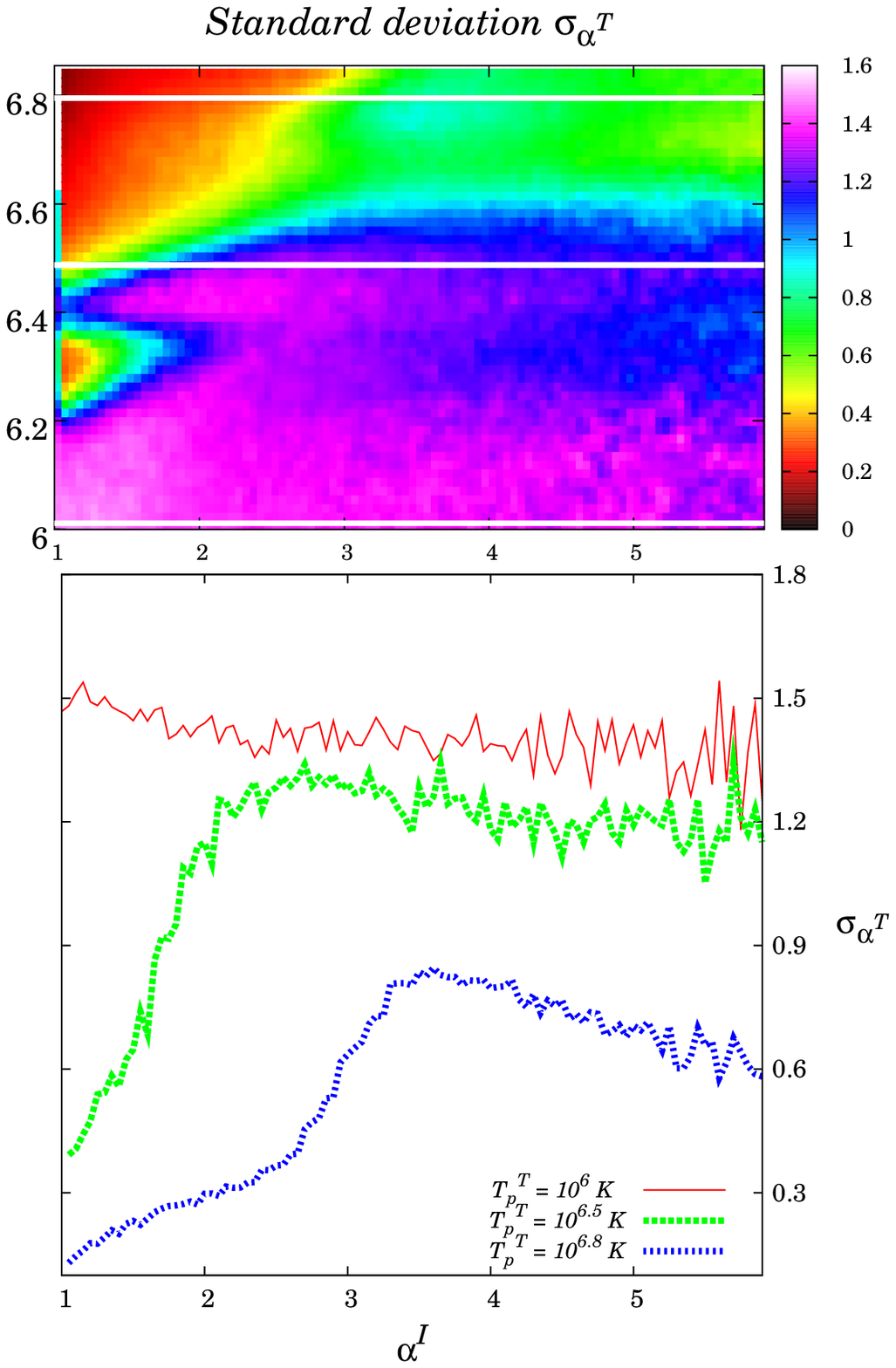}%
\end{center}
\caption{Same as Figure~\ref{fig:statistical}, but considering now the second set of uncertainties (leading to a total uncertainty ranging between 25 and 30\%, see Section~\ref{sec:uncertainties}). For AR DEMs with high temperature peak, the confidence level is significantly decreased from 0.9 to 0.6. However, for low temperature peaks AR DEMs, results are similar. \label{fig:statistical2}}
\end{figure*}

The summarized results regarding the second set of uncertainties used in this work and described in Section~\ref{sec:uncertainties} is displayed in Figure~\ref{fig:statistical2}. In this case, the atomic physics uncertainties are greatly reduced from 20\% to 10\% for class 1 and from 30\% to 10\% for classes 2 through 4, while the calibration uncertainties are reduced from 25\% to 20\%. The resulting total uncertainty varies between 25-30\% depending on the line. As expected, the reduced uncertainties lead to an improved correlation between the estmated slope $\alpha^I$ and the true one $\alpha^T$, particularly for medium temperature peak around $10^{6.5}$~K. As a result, the standard deviation is decreased, ranging now between 0.2 and 0.8 for $T_p^T = 10^{6.8}$~K, 0.3 and 1.2 for $T_p^T=10^{6.5}$~K, and approaching the same constant value as before, around $\sigma_{\alpha^T} = 1.4$. Maps like these in Figures~\ref{fig:statistical} and~\ref{fig:statistical2} are useful for interpreting the DEM inversions from true observations: given the slope and the temperature of the peak, both mean value and confidence level can be derived.

The confidence levels derived in the present work can be used to evaluate the agreement between theoretical model predictions and DEM
measurements. In the recent paper of \citet{bradshaw2012}, the authors carried out a series of low-frequency nanoflare simulations. They investigated a large number of heating and coronal loop properties, such as the magnitude and duration of the nanoflares and the length of the loop. They concluded that the low frequency heating mechanism cannot explain DEM slopes $\alpha \geq 2.6$, similar to the findings of \citet{mulu2011}. Comparing their results to the current observations of AR cores (see Section~\ref{sec:motivations} for corresponding references), they found that 36\% of observed AR cores are consistent with low-frequency nanoflare heating if uncertainties in the slope
measurements are ignored. Using then the slope uncertainties estimated around $\Delta \alpha \pm 1$ in this work, they concluded that as few as zero to as many as 77\% of AR cores are actually consistent with low-frequency nanoflares. More recently, \citet{reep2013} studied a scenario they call a "nanoflare train" in which a finite series of high-frequency nanoflares occur within the same loop strand and then cease. The predicted slopes are in the range $0.88 \leq \alpha \leq 4.56$. Using again an uncertainty of $\Delta \alpha \pm 1$, they concluded that 86\% to 100\% of current AR core observations are consistent with such trains.

The determination of the uncertainties associated with the atomic physic processes is no simple matter, as discussed in Section~\ref{sec:uncertainties}, that is why we have tested two different sets of uncertainties. However, the most important issue here, considering the temperature peaks currently derived in observational analysis, is that whatever the set of uncertainties used to determine the confidence level on the reconstructed slope, their typical values remain important relative to what is necessary to strongly constrain the timescale of the coronal heating. \citet{warren2012} and \citet{winebarger2012} for example, derived temperature peak generally around $\log T_e = 6.6$, whereas \citet{schmelz2012} derived temperature peaks generally between $\log T_e = 6.5$ and $\log T_e = 6.7$. For these typical values, the slope uncertainties varies between $\Delta \alpha = \pm 0.9$ and 1.3 for slopes $\alpha > 3$ when using the first set of uncertainties, and it varies between $\pm 0.6$ and 1.0 when using the second set of smaller uncertainties.  It appears, therefore, that it is not yet possible to place strong constraints on the coronal heating timescale using observed DEM slopes and the predictions of theoretical models. Further improvements in reducing atomic physics uncertainties are highly desirable.


\section{Summary and conclusions}
\label{sec:conclusions}
The slope of the DEM distribution coolward of the coronal peak can potentially be used to diagnose the timescale of energy deposition in the solar corona. Indeed the DEM slope provides important information on the proportion of hot to warm material, which is useful to determine the heating timescale. Recent observational studies of AR cores suggest that some active region cores are consistent with low frequency heating mechanisms, where the plasma cools completely before being reheated, while other show consistency with high frequency energy deposition, where rapid reheating causes the temperature to fluctuate about a particular value. Distinguishing between these possibilities is important for identifying the physical mechanism of the heating. It is therefore crucial to understand the uncertainties in measurements of observed DEM slopes.

In this work, we presented an application of our recently developed technique in the specific case of typical AR DEMs, in order to properly estimate confidence level of the observed DEM slopes and assist the DEM interpretation. Using a probabilistic approach and Monte-Carlo simulations of uncertainties to interpret the DEM inversion, our method is useful for examining the robustness of the DEM inversion, and to analyze the DEM inversion properties. Comparing simulated observations of the \textit{Hinode}/EIS spectrometer with inferred results, the range or multiple ranges of solutions consistent with a given set of measurement can be estimated, along with their associated probabilities. From such probability distributions, statistical quantities can be derived, such as the standard deviation, providing rigorous confidence levels on the DEM solutions.

In this way, we carefully assess the errors in the DEM slopes determined from \textit{Hinode}/EIS data. Both random and systematic errors have been taken into account. We paid particular attention to the description of the systematic errors related to the atomic physics process and abundances. Uncertainties associated with ionization fractions, elemental abundances, FIP effect and a combination of uncertainties in the radiative and excitation rates have been simulated. Additional systematic errors have been added on the blended lines, to take into account the technical difficulties in isolating a single line intensity. We argue that our work actually provides an optimistic estimation of the slope confidence levels: the mathematical difficulties intrinsic to solving an inverse problem introduce additional ambiguity, while our method allows to focus only on the impact of intrinsic uncertainties. The fact that our inverted DEMs have the same functional form than the true ones, known \textit{a priori}, means that our slope uncertainties are lower limits. In reality, the form of the true DEM is unknown, and this introduces additional uncertainty, through the use of blind inversion.    

In Section~\ref{sec:results}, we demonstrated how the slope reconstruction is affected by the uncertainties. The analysis of the probability maps provides the range of slopes consistent with the observed DEM slopes. These maps show that in most cases, a large range of solutions is consistent with the measurements. The presence of uncertainties degrades the quality of the inversion, leading to typical confidence levels around 0.9-1.0. However, the inversion robustness, and thus the confidence level, largely depends on the number of lines constraining the slope. For DEMs with high temperature peaks [5-6~MK], about 20 lines contain suitable information, while low temperature peaks [1-3~MK] reduce this number to less than 10. For these latter cases, the effect of uncertainties leads to larger confidence levels, about 1.3 and more in some cases. 

The slope confidence levels derived in the present work are useful for quantifying the degree of agreement between theoretical models and observations. Current slope reconstructions can thus be properly compared to theoretical expectations. However, the typical derived confidence levels remain significant, comparing to the majority of observed slopes values concentrated between 1.5 and 5. The sizable confidence levels make it difficult to draw definitive conclusions about the suitability of a given heating model, implying in one hand, that a model might be consistent with the majority of observations or, in the other hand, with none at all~\citep[see][for a practical application of these confidence levels]{bradshaw2012}. When relaxing the constraint on the DEM slopes as \citet{reep2013}, the slope DEM diagnostic does not allow to distinguish between different scenarios, because observations can thus be explained by a variety of different heating models. 

Our generic approach can be improved for specific datasets, and additional sophistication in can be incorporated (see Section~\ref{sec:uncertainties}). We could, for example, use a customized set of uncertainties for a given set of lines. However, the main important point of our work, is that, even for uncertainties that would seem to be on the low end of what is feasible (our second set of uncertainties), the corresponding uncertainty in the measured slope may be too large to definitively exclude or corroborate a given heating scenario in many cases. The methodology presented here can also be used to establish the optimal set of lines required to obtain the smallest possible confidence levels. Such kind of preliminary investigations can be very helpful to optimize the future instruments, whether it be spectrometer or broad band imagers, in order to maximize their DEM diagnostic capabilities.      

\acknowledgments
S.P. acknowledges the support from the Belgian Federal Science Policy Office through the international cooperation programmes and the ESA-PRODEX programme and the support of the Institut d'Astrophysique Spatiale (IAS). F.A. acknowledges the support of the Royal Observatory of Belgium. The work of J.A.K. was supported by the NASA Supporting Research and Technology Program. The authors would like to thank G. Del Zanna, H. Warren, G. Doschek, M. Laming, E. Landi, H. Mason, J. Schmelz and P. Young for fruitful discussions and comments about atomic physic uncertainties. Discussions with H. Mason, H. Warren, and P. Testa at the second meeting of the Bradshaw/Mason International Space Science Institute Team were also very helpful.

\clearpage

\begin{table}[p]
\centering
\begin{tabular}{lccc}

Ions &	Wavelength ($\mathring{\rm A}$) &	$\log (T [K])$ & Total uncertainty $\sigma_{unc}$\\
\hline
\hline
Mg V	& 276.579 & 5.45 & 61.03 \% \\
Mg VI & 268.991 & 5.65 & 61.03 \% \\
Mg VI & 270.391 & 5.65 & 61.03 \% \\
Mg VII$^\textit{b}$ & 278.404 & 5.80 & 62.85 \% \\
Mg VII & 280.745 & 5.80 & 61.03 \% \\
\hline
Si VII & 275.354 & 5.80 & 61.03 \% \\
Si IX & 258.082 & 6.05 & 61.03 \% \\
Si X & 258.371 & 6.15 & 61.03 \% \\
Si X & 261.044 & 6.15 & 61.03 \% \\
\hline
Fe IX & 188.497 & 5.85 & 61.03 \% \\
Fe IX & 197.865 & 5.85 & 61.03 \% \\
Fe X & 184.357 & 6.05 & 61.03 \% \\
Fe XI & 180.408 & 6.15 & 61.03 \% \\
Fe XI & 188.232 & 6.15 & 61.03 \% \\
Fe XII & 192.394 & 6.20 & 61.03 \% \\
Fe XII & 195.119 & 6.20 & 61.03 \% \\
Fe XIII & 202.044 & 6.25 & 61.03 \% \\
Fe XIII & 203.828 & 6.25 & 61.03 \% \\
Fe XIV & 264.790 & 6.30 & 61.03 \% \\
Fe XIV & 270.522 & 6.30 & 61.03 \% \\
Fe XIV$^\textit{b}$ & 274.204 & 6.30 & 62.85 \% \\
Fe XV & 284.163 & 6.35 & 61.03 \%\\
Fe XVI & 262.976 & 6.45 & 61.03 \% \\
\hline
S X & 264.231 & 6.15 & 53.15 \% \\
S XIII$^\textit{b}$ & 256.685 & 6.40 & 55.23 \% \\
\hline
Ca XIV & 193.866 & 6.55 & 61.03 \% \\
Ca XV & 200.972 & 6.65 & 61.03 \% \\
Ca XVI & 208.604 & 6.70 & 61.03 \% \\
Ca XVII$^\textit{b}$ & 192.853 & 6.75 & 62.85 \% \\

\end{tabular}
\caption{List of the \textit{Hinode}/EIS spectral lines used in our simulations. Lines are sorted by elements as a function of the peak temperature of the contribution functions. The blended lines are specified with the index $b$. The fourth column indicate the percentage of total uncertainty applied to each spectral lines, resulting of both systematic and random errors.\label{tab:eis_lines}}
\end{table}

\clearpage

\clearpage

\clearpage

\clearpage

\clearpage

\clearpage

\clearpage

\end{document}